\documentclass[
  a4paper,
  onecolumn,
  11pt,
  allowtoday
]{quantumarticle}
\pdfoutput=1

\usepackage[utf8]{inputenc}
\usepackage[english]{babel}
\usepackage[T1]{fontenc}

\usepackage{amsmath,amssymb}
\usepackage{natbib}
\usepackage{hyperref}
\usepackage{graphicx}
\usepackage{float}
\usepackage{xcolor}
\usepackage{tikz}
\usepackage{array}
\usepackage{algorithm}
\usepackage{algpseudocode}
\usepackage{booktabs}

\usepackage{tikzit}

\tikzstyle{gate}=[shape=rectangle, text height=1.5ex, text depth=0.25ex, yshift=0.5mm, fill=white, draw=black, minimum height=5mm, yshift=-0.5mm, minimum width=5mm, font={\small}, tikzit category=circuit]
\tikzstyle{big gate}=[shape=rectangle, text height=1.5ex, text depth=0.25ex, yshift=0.5mm, fill=white, draw=black, minimum height=10mm, yshift=-0.5mm, minimum width=5mm, font={\small}, tikzit category=circuit]
\tikzstyle{Z dot}=[inner sep=0mm, minimum size=2mm, shape=circle, draw=black, fill={rgb,255: red,216; green,248; blue,216}, tikzit category=zx]
\tikzstyle{Z dot bb}=[inner sep=0mm, minimum size=2mm, shape=circle, draw=black, fill={rgb,255: red,216; green,248; blue,216}, line width=1.1pt, tikzit category=zx]
\tikzstyle{Z phase dot}=[minimum size=5mm, font={\footnotesize\boldmath}, shape=rectangle, rounded corners=2mm, inner sep=0.2mm, outer sep=-2mm, scale=0.8, tikzit shape=rectangle, draw=black, fill={rgb,255: red,216; green,248; blue,216}, tikzit draw=blue, tikzit category=zx]
\tikzstyle{Z phase dot bb}=[minimum size=5mm, font={\footnotesize\boldmath}, shape=rectangle, rounded corners=2mm, inner sep=0.2mm, outer sep=-2mm, scale=0.8, tikzit shape=rectangle, draw=black, fill={rgb,255: red,216; green,248; blue,216}, tikzit draw=blue, line width=1.1pt, tikzit category=zx]
\tikzstyle{X dot}=[Z dot, shape=circle, draw=black, fill={rgb,255: red,232; green,165; blue,165}, tikzit category=zx]
\tikzstyle{X dot bb}=[inner sep=0mm, minimum size=2mm, shape=circle, draw=black, fill={rgb,255: red,232; green,165; blue,165}, line width=1.1pt, tikzit category=zx]
\tikzstyle{X phase dot}=[Z phase dot, tikzit shape=rectangle, tikzit draw=blue, fill={rgb,255: red,232; green,165; blue,165}, font={\footnotesize\boldmath}, tikzit category=zx]
\tikzstyle{X phase dot bb}=[minimum size=5mm, font={\footnotesize\boldmath}, shape=rectangle, rounded corners=2mm, inner sep=0.2mm, outer sep=-2mm, scale=0.8, tikzit shape=rectangle, draw=black, fill={rgb,255: red,232; green,165; blue,165}, tikzit draw=blue, line width=1.1pt, tikzit category=zx]
\tikzstyle{hadamard}=[fill=yellow, draw=black, shape=rectangle, inner sep=0.6mm, minimum height=1.5mm, minimum width=1.5mm, tikzit category=zx]
\tikzstyle{paulibox}=[fill={rgb,255: red,221; green,221; blue,255}, draw=black, shape=rectangle, inner sep=0.6mm, minimum height=5mm, minimum width=5mm, font={\footnotesize}, text height=1.5ex, text depth=0.25ex, tikzit category=zx]
\tikzstyle{vertex}=[inner sep=0mm, minimum size=1mm, shape=circle, draw=black, fill=black, tikzit category=misc]
\tikzstyle{vertex set}=[inner sep=0mm, minimum size=1mm, shape=circle, draw=black, fill=white, font={\footnotesize\boldmath}, tikzit category=misc]
\tikzstyle{small black dot}=[fill=black, draw=black, shape=circle, inner sep=0pt, minimum width=1.2mm, tikzit category=circuit]
\tikzstyle{cnot ctrl}=[fill=black, draw=black, shape=circle, inner sep=0pt, minimum width=1.2mm, tikzit category=circuit]
\tikzstyle{cnot targ}=[fill=white, draw=white, shape=circle, tikzit category=circuit, label={center:$\oplus$}, inner sep=0pt, minimum width=2.1mm, tikzit fill={rgb,255: red,102; green,204; blue,255}, tikzit draw=black]
\tikzstyle{ket}=[fill=white, draw=black, shape=regular polygon, regular polygon sides=3, regular polygon rotate=-30, scale=0.7, inner sep=1pt, tikzit category=circuit, tikzit shape=rectangle, tikzit fill=green]
\tikzstyle{bra}=[fill=white, draw=black, shape=regular polygon, regular polygon sides=3, regular polygon rotate=30, scale=0.7, inner sep=1pt, tikzit category=circuit, tikzit shape=rectangle, tikzit fill=red]
\tikzstyle{scalar}=[shape=rectangle, text height=1.5ex, text depth=0.25ex, yshift=0.5mm, fill=white, draw=black, minimum height=5mm, yshift=-0.5mm, minimum width=5mm, font={\small}, rounded corners=2mm]
\tikzstyle{clabel}=[fill=white, draw=none, shape=rectangle, tikzit fill={rgb,255: red,56; green,255; blue,242}, font={\footnotesize}, inner sep=1pt, tikzit category=labels]
\tikzstyle{empty diagram}=[draw={gray!40!white}, dashed, shape=rectangle, minimum width=1cm, minimum height=1cm, tikzit category=misc]
\tikzstyle{amap}=[fill=white, draw=black, shape=NEbox, tikzit category=asymmetric, tikzit fill=yellow, tikzit shape=rectangle]
\tikzstyle{amap conj}=[fill=white, draw=black, shape=NWbox, tikzit category=asymmetric, tikzit fill=green, tikzit shape=rectangle]
\tikzstyle{amap adj}=[fill=white, draw=black, shape=SEbox, tikzit category=asymmetric, tikzit fill=red, tikzit shape=rectangle]
\tikzstyle{amap trans}=[fill=white, draw=black, shape=SWbox, tikzit category=asymmetric, tikzit fill=orange, tikzit shape=rectangle]
\tikzstyle{astate}=[fill=white, draw=black, shape=NEtriangle, tikzit category=asymmetric, tikzit shape=circle, tikzit fill=yellow]
\tikzstyle{astate conj}=[fill=white, draw=black, shape=NWtriangle, tikzit category=asymmetric, tikzit shape=circle, tikzit fill=green]
\tikzstyle{astate adj}=[fill=white, draw=black, shape=SEtriangle, tikzit category=asymmetric, tikzit shape=circle, tikzit fill=red]
\tikzstyle{astate trans}=[fill=white, draw=black, shape=SWtriangle, tikzit category=asymmetric, tikzit shape=circle, tikzit fill=orange]
\tikzstyle{bigbox}=[fill=white, draw=black, shape=rectangle, tikzit category=zx, minimum width=1.5cm, minimum height=2.6cm]

\tikzstyle{hadamard edge}=[-, dashed, dash pattern=on 2pt off 0.5pt, thick, draw={rgb,255: red,68; green,136; blue,255}]
\tikzstyle{box edge}=[-, dashed, dash pattern=on 2pt off 0.5pt, thick, draw={rgb,255: red,203; green,192; blue,225}]
\tikzstyle{brace edge}=[-, tikzit draw=blue, decorate, decoration={brace,amplitude=1mm,raise=-1mm}]
\tikzstyle{diredge}=[->]
\tikzstyle{double edge}=[-, double, shorten <=-1mm, shorten >=-1mm, double distance=2pt]
\tikzstyle{gray edge}=[-, {gray!60!white}]
\tikzstyle{pointer edge}=[->, very thick, gray]
\tikzstyle{boldedge}=[-, line width=1.1pt, shorten <=-0.17mm, shorten >=-0.17mm]
\tikzstyle{bidir edge}=[<->, very thick, draw={rgb,255: red,191; green,191; blue,191}]
\tikzstyle{separator edge}=[-, dashed, dash pattern=on 2pt off 0.5pt, thick, draw={rgb,255: red,153; green,153; blue,153}]
\tikzstyle{dashed edge}=[-, dashed, dash pattern=on 2pt off 0.5pt, thick]
\tikzstyle{solid edge}=[-, thick]

\makeatletter
\providecommand\caption@documentclass{standard}
\providecommand\@afterenddocumenthook{}
\makeatother

\usepackage{listings}
\definecolor{codebg}{HTML}{F7F7F7}
\definecolor{codeframe}{HTML}{CCCCCC}
\definecolor{codekw}{HTML}{1A56DB}
\definecolor{codecomment}{HTML}{6B7280}
\definecolor{codestr}{HTML}{9D174D}
\newcommand{\bigO}[1]{\mathcal{O}(#1)}
\newcommand{\FF}{\mathrm{GF}(2)}
\newcommand{\RR}{\mathbb{R}}
\newcommand{\ngates}{G}          
\newcommand{\nmagic}{T}          
\newcommand{\nerr}{E}            
\newcommand{\navgflip}{\delta}   
\newcommand{\perr}{p}            
\newcommand{\batchsize}{s}       
\newcommand{\stabrank}{\chi}     
\newcommand{\decomprate}{\alpha} 
\newcommand{\numqubits}{n}      

\begin{document}

\title{Tsim: Fast Universal Simulator for Quantum Error Correction}
\date{April 1, 2026}

\author{Rafael Haenel}
\email{rhaenel@quera.com}
\affiliation{QuEra Computing Inc., Boston, USA}

\author{Xiuzhe Luo}
\affiliation{QuEra Computing Inc., Boston, USA}

\author{Chen Zhao}
\email{czhao@quera.com}
\affiliation{QuEra Computing Inc., Boston, USA}

\maketitle

\begin{abstract}
  We present Tsim, an open-source high-throughput simulator for universal noisy quantum circuits targeting quantum error correction. Tsim represents quantum circuits as ZX diagrams, where Pauli channels are modeled as parameterized vertices. Diagrams are simplified via parameterized ZX rules, and then compiled for vectorized sampling with GPU acceleration. After the one-time compilation, one can sample detector or measurement shots in linear time in the number of Clifford gates and exponentially only in the number of non-Clifford gates. Tsim implements the Stim API and fully supports the Stim circuit format, extending it with T and arbitrary single-qubit rotation instructions. For low-magic circuits, Tsim throughput can match the sampling performance of Stim.
\end{abstract}

\section{Introduction}

\begin{figure}[t]
  \centering
  \includegraphics[width=1.0\linewidth]{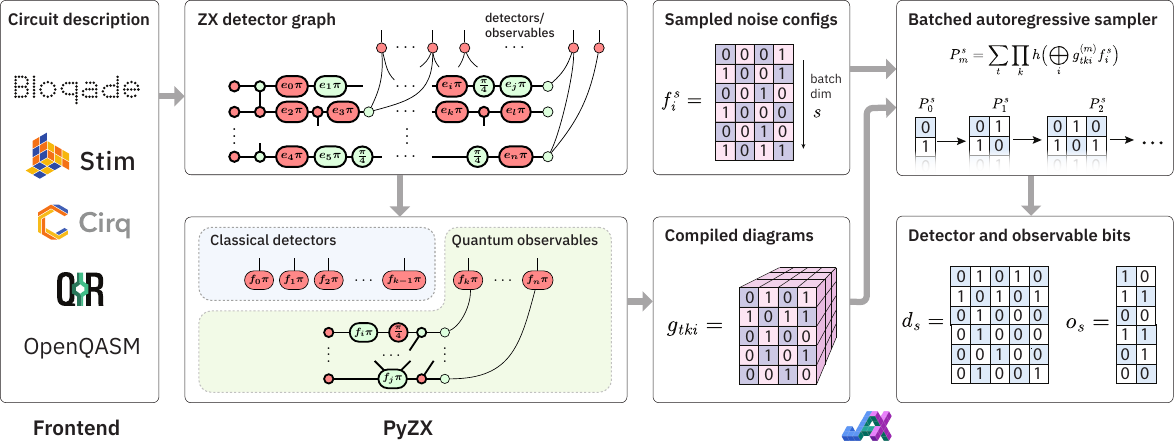}
  \caption{Overview of the Tsim simulation pipeline.
    Quantum programs are translated into ZX diagrams in which Pauli noise channels appear as parameterized vertices with binary variables $e_i$.
    ZX simplification factors the diagram into a classical part, that represents the Tanner graph, and a quantum part containing the observable circuit. Both parts define a new basis of \textit{error mechanisms} $f_i = \bigoplus_j T_{ij}\,e_j$.
    The observable diagram
    is used to compute marginal probabilities for autoregressive sampling. Here, each diagram is decomposed into a sum of Clifford terms via stabilizer rank decomposition~\cite{Sutcliffe_2025} and compiled into binary JAX tensors $g_{tki}$.
    At sampling time, JIT-compiled XLA kernels contract $g_{tki}$ with batched noise configurations $f_i^{s}$ to evaluate marginal probabilities and autoregressively sample detector and observable bits.
  }
  \label{fig:pipeline}
\end{figure}

Quantum computing at utility scale requires quantum error correction (QEC) to achieve sufficiently low logical error rates. In recent years, quantum hardware has made significant strides toward this goal, with experimental demonstrations of key fault-tolerant primitives \cite{google2023suppressing, ryan2024high, sales2025experimental, rosenfeld2025magic, bluvstein2026fault}. Despite these advances, large-scale fault-tolerant quantum computers remain beyond current hardware capabilities. Consequently, the development, benchmarking, and validation of QEC protocols continue to rely heavily on classical simulation tools. To address this need, we introduce Tsim, a GPU-accelerated universal quantum circuit simulator based on stabilizer rank decomposition and the ZX-calculus. Tsim addresses the following key requirements of a practical QEC simulator.

First, QEC simulations are typically performed at the gadget level. Rather than simulating full logical algorithms, one focuses on individual components such as stabilizer extraction circuits (e.g., quantum memory), single logical operations (e.g., lattice surgery and transversal gates \cite{horsman2012surface, zhou2024algorithmic}), or magic-state preparation protocols (e.g., magic state cultivation \cite{gidney2024magic}). Such gadgets can involve hundreds to thousands of physical qubits, yet are predominantly---and sometimes entirely---composed of Clifford operations, reflecting the structure of stabilizer codes. A practical QEC simulator must therefore scale to large qubit counts while exploiting the low-magic structure of these circuits.
Tsim supports low-magic circuit simulation via stabilizer rank decomposition, yielding runtimes that scale polynomially in the number of qubits and exponentially only in the non-Clifford gate count.

Second, QEC protocols must be evaluated under realistic noise models. In practice, much of the community's simulation effort focuses on Pauli noise at the circuit level~\cite{derks2025designing}. This approximation is often well justified: stabilizer measurements effectively project general noise channels onto discrete bit and phase flips. As a result, an efficient QEC simulator must handle Pauli errors with minimal overhead.
Tsim addresses this by representing Pauli errors as parameterized ZX vertices, so that the expensive diagrammatic reduction needs only be performed once and is reused across noise configurations~\cite{Sutcliffe_2025}. This allows Tsim to reach linear-time sampling as in Stim for Clifford circuits with deterministic detectors.

Third, simulators demand extremely high throughput to support Monte Carlo sampling at low error rates \cite{gidney2021stim}. For example, estimating the logical error rate of a QEC gadget at $10^{-9}$--$10^{-12}$ requires generating billions to trillions of shots from the probability distribution of its quantum circuit. Consequently, an ideal QEC simulator must support extremely high-throughput sampling. Tsim leverages JAX~\cite{jax2018github} and XLA compilation to vectorize shot execution on CPU and GPU, enabling high-throughput Monte Carlo sampling.

In the landscape of existing QEC simulators, Stim~\cite{gidney2021stim} has emerged as the de facto standard for protocols composed entirely of Clifford gates and Pauli errors, achieving linear-time sampling via Pauli frame tracking and high throughput through vectorized execution. However, many practically relevant QEC gadgets contain a small number of non-Clifford operations, which require simulation techniques beyond the stabilizer formalism. Existing approaches include state-vector simulators (exponential in qubit count), tensor network methods such as matrix product states (efficient for limited entanglement), and low-magic simulation techniques such as extended tableau methods~\cite{yoder2012generalization,li2025soft}, Pauli propagation~\cite{rall2019simulation,rudolph2025pauli}, and stabilizer rank decomposition~\cite{bravyi2019simulation}. Among these, stabilizer rank decomposition combined with ZX-calculus simplification~\cite{kissinger2022classical,kissinger2022simulating,Sutcliffe_2025,sutcliffe2025novel,sutcliffe2024procedurally,codsi2022cutting} is particularly well suited to the high-entanglement, low-magic structure of QEC circuits, and has been shown to exploit protocol structure to achieve significantly lower stabilizer rank than previously known~\cite{wan2025cutting,wan2026simulatingmagicstatecultivation}. Tsim builds on these advances and is implemented on top of PyZX~\cite{kissinger2020Pyzx}.

Tsim implements the Stim API for seamless integration with existing workflows and extends the Stim instruction set with non-Clifford gates. It supports sampling of either measurement outcomes or detector events. The software is available open-source under the Apache-2.0 license\footnote{\url{https://github.com/QuEraComputing/tsim}}. Tsim places strong emphasis on reliability and includes an automated test suite with more than 95\% coverage.

This manuscript is organized as follows. In Sec.~\ref{sec:pipeline}, we present an overview of Tsim and describe the simulation pipeline in detail. In Sec.~\ref{sec:benchmarks}, we present performance benchmarks. We conclude in Sec.~\ref{sec:summary}. Additionally, we provide a summary of the ZX-calculus, a more detailed description of the simulation strategy, and the Tsim API in the Appendix.


\section{Methods}
\label{sec:pipeline}

Tsim is distributed as a Python package (\texttt{bloqade-tsim}) and mirrors the full Stim Python API~\cite{gidney2021stim}. All Stim features are supported, including Clifford gates, single-qubit measurements and resets, Pauli product measurements, classically controlled Pauli gates, correlated Pauli noise channels, and detector and observable annotations. Tsim extends the instruction set with non-Clifford gates \texttt{T}, \texttt{T\_DAG}, \texttt{R\_X}, \texttt{R\_Y}, \texttt{R\_Z}, and \texttt{U3}. For many existing Stim-based workflows, a good starting point is to replace \texttt{import stim} with \texttt{import tsim}.

The following listing illustrates Tsim on a minimal noisy non-Clifford circuit. Additionally, Tsim is tightly integrated into the Bloqade framework \cite{bloqade2025} (see Appendix~\ref{sec:frontend}).

\begin{lstlisting}[caption={Minimal Tsim detector-sampling example.}, label={lst:detector_sampling}]
  import tsim

  c = tsim.Circuit("""
    RX 0
    R 1
    R_Z(0.125) 0  # 0.125*pi rotation around Z
    PAULI_CHANNEL_1(0.1, 0.1, 0.2) 0 1
    H 0
    CNOT 0 1
    DEPOLARIZE2(0.01) 0 1
    M 0 1
    DETECTOR rec[-1] rec[-2]
  """)

  sampler = c.compile_detector_sampler()
  samples = sampler.sample(shots=1_000_000)
\end{lstlisting}

The simulation workflow follows a compile-once, sample-many pattern (Fig.~\ref{fig:pipeline}). Tsim accepts quantum programs specified in Stim,
Bloqade~\cite{bloqade2025},
OpenQASM~\cite{cross2017open},
Cirq~\cite{cirq2025},
and QIR (Quantum Intermediate Representation)~\cite{qir2026spec}; these front-end formats are translated to Tsim's intermediate representation (IR).

The noisy circuit in the IR is converted into a parameterized ZX diagram, where Pauli noise channels enter as symbolic parameters rather than concrete gates. Diagrammatic reduction rules (via PyZX) then simplify the parameterized diagram, usually resulting in disconnected detector and observable components. The reduced diagram is assembled into a contiguous array data structure, the compiled sampler, which is reused across noise configurations and sampling calls.

At sampling time, the compiled sampler draws shots by evaluating the reduced diagram for many independent noise realizations in parallel, accelerated by JAX/XLA~\cite{jax2018github} on CPU or GPU. Circuits can be compiled into either measurement samplers or detector samplers. Measurement samplers return raw measurement outcomes, whereas detector samplers return detector and logical observable bits. As we show in Sec.~\ref{sec:detector_advantage}, detector sampling often admits a more efficient ZX formulation, since ZX reduction can separate detector bits from the logical circuit and evaluate each detector as an independent connected component.

\subsection{ZX representation}
Internally, Tsim represents circuits as ZX diagrams using the \textit{doubled} notation of the ZX-calculus for more compact representation of measurements (see Appendix~\ref{sec:doubling}). In this notation, bold wires and spiders denote quantum objects, while thin wires and spiders carry classical information such as measurement outcomes. Measurements correspond to thin spiders that consume a quantum wire and produce a classical one. Classical wires support non-unitary operations such as COPY and XOR, which allow native representation of detectors and observables within the diagram: detectors and observables are defined as XOR combinations of measurement results, and measurement results need to be copied because they may appear in multiple detectors and observables.
\begin{equation}
  \begin{tikzpicture}
	\begin{pgfonlayer}{nodelayer}
		\node [style=Z dot] (16) at (-10.25, -0.25) {};
		\node [style=none] (184) at (-11.5, -0.25) {};
		\node [style=none] (189) at (-9.5, 0.5) {};
		\node [style=Z dot bb] (190) at (-19.5, 0.25) {};
		\node [style=none] (191) at (-20.75, 0.25) {};
		\node [style=none] (192) at (-18.25, 0.25) {};
		\node [style=X dot bb] (193) at (-19.5, -0.75) {};
		\node [style=none] (194) at (-20.75, -0.75) {};
		\node [style=none] (195) at (-18.25, -0.75) {};
		\node [style=Z dot] (201) at (-4, -0.25) {};
		\node [style=none] (202) at (-5, -0.25) {};
		\node [style=none] (203) at (-3, 0.25) {};
		\node [style=none] (204) at (-3, -0.5) {};
		\node [style=none] (205) at (-22.5, -0.25) {CNOT};
		\node [style=none] (206) at (-14, -0.25) {MEASURE};
		\node [style=none] (207) at (-6.5, -0.25) {COPY};
		\node [style=X dot] (209) at (2.75, -0.25) {};
		\node [style=none] (210) at (1.75, 0) {};
		\node [style=none] (211) at (3.75, -0.25) {};
		\node [style=none] (212) at (1.75, -0.5) {};
		\node [style=none] (213) at (0.5, -0.25) {XOR};
	\end{pgfonlayer}
	\begin{pgfonlayer}{edgelayer}
		\draw [style=boldedge] (184.center) to (16);
		\draw [bend left=15] (16) to (189.center);
		\draw [style=boldedge] (191.center) to (190);
		\draw [style=boldedge] (190) to (192.center);
		\draw [style=boldedge] (194.center) to (193);
		\draw [style=boldedge] (193) to (195.center);
		\draw [style=boldedge] (190) to (193);
		\draw [style=boldedge] (194.center) to (194.center);
		\draw (202.center) to (201);
		\draw [bend left=15] (201) to (203.center);
		\draw [bend right=15] (201) to (204.center);
		\draw [bend left=15] (210.center) to (209);
		\draw (209) to (211.center);
		\draw [looseness=1.25, bend left=15] (209) to (212.center);
	\end{pgfonlayer}
\end{tikzpicture}
  \label{eqn:doubled_notation}
\end{equation}
Doubled diagrams can easily be transpiled into regular diagrams following Eqs.~\eqref{eqn:doubling}--\eqref{eqn:doubling2}.

Pauli noise channels are represented as parameterized X and Z spiders with phase in $[0,\pi]$.
For example, X and Z bit-flip channels are represented by the following ZX diagrams:
\begin{equation}
  \begin{tikzpicture}
	\begin{pgfonlayer}{nodelayer}
		\node [style=X phase dot bb] (1) at (-4.25, -0.25) {$\;e_i\pi\;$};
		\node [style=none] (3) at (-5.5, -0.25) {};
		\node [style=none] (4) at (-3, -0.25) {};
		\node [style=Z phase dot bb] (12) at (5.75, -0.25) {$\;e_i\pi\;$};
		\node [style=none] (13) at (4.5, -0.25) {};
		\node [style=none] (14) at (7, -0.25) {};
		\node [style=none] (15) at (-8.5, -0.25) {X flip channel};
		\node [style=none] (16) at (1.5, -0.25) {Z flip channel};
	\end{pgfonlayer}
	\begin{pgfonlayer}{edgelayer}
		\draw [style=boldedge] (3.center) to (1);
		\draw [style=boldedge] (12) to (14.center);
		\draw [style=boldedge] (1) to (4.center);
		\draw [style=boldedge] (13.center) to (12);
	\end{pgfonlayer}
\end{tikzpicture}
\end{equation}
In this way, Tsim supports arbitrary correlated Pauli channels as detailed in Appendix~\ref{sec:pauli_channels}.
The parameterized representation of Pauli channels allows ZX reduction and compilation in a way that
is agnostic to a specific noise configuration.

\subsection{Simulation pipeline}

We describe the simulation pipeline using an example circuit containing two distance-2 repetition code blocks. The circuit is structured as follows: First, a round of stabilizer readout is performed, with CNOT gates followed by bit flip error channels. Next, we apply two logical $\pi/8$ rotations, separated by a phase flip error channel. Then, a transversal CNOT gate is applied between the two code blocks. Finally, a second round of stabilizer readout is performed, followed by terminal measurements. We define standard detectors as XOR operations
that check the parity of data qubits between subsequent rounds (see Sec.~\ref{sec:detectors_observables}). Two observables
are defined from the terminal measurements. This yields the following diagram:
\begin{equation}
  \begin{tikzpicture}
	\begin{pgfonlayer}{nodelayer}
		\node [style=X dot bb] (38) at (9.75, -4.5) {};
		\node [style=none] (77) at (17.25, 2.5) {};
		\node [style=none] (73) at (3.5, 2.5) {};
		\node [style=none] (74) at (10, 2.5) {};
		\node [style=none] (76) at (16.5, 2.5) {};
		\node [style=X dot bb] (49) at (13, -4.5) {};
		\node [style=X phase dot bb] (52) at (14.25, -5.5) {$\;e_{16}\pi\;$};
		\node [style=Z dot] (46) at (15.5, 0.75) {};
		\node [style=X phase dot bb] (22) at (-1, -3.5) {$\;e_2\pi\;$};
		\node [style=none] (79) at (19.5, 2.5) {};
		\node [style=X dot bb] (27) at (0.25, -4.5) {};
		\node [style=X dot bb] (25) at (0.25, -0.25) {};
		\node [style=X dot bb] (10) at (-2.25, -4.5) {};
		\node [style=X dot] (70) at (18.75, 1.75) {};
		\node [style=X phase dot bb] (29) at (1.5, -0.25) {$\;e_4\pi\;$};
		\node [style=Z dot bb] (17) at (-2.25, -3.5) {};
		\node [style=X phase dot bb] (50) at (14.25, -1.25) {$\;e_{14}\pi\;$};
		\node [style=Z dot bb] (24) at (10.5, 0.75) {};
		\node [style=X phase dot bb] (20) at (-1, 0.75) {$\;e_0\pi\;$};
		\node [style=X phase dot bb] (43) at (11.75, -0.25) {$\;e_{10}\pi\;$};
		\node [style=Z dot bb] (18) at (0.25, -1.25) {};
		\node [style=X phase dot bb] (45) at (11.75, -4.5) {$\;e_{12}\pi\;$};
		\node [style=X phase dot bb] (28) at (1.5, -1.25) {$\;e_5\pi\;$};
		\node [style=X dot bb] (4) at (-2.25, -0.25) {};
		\node [style=X dot bb] (9) at (-3.25, -4.5) {};
		\node [style=Z dot] (35) at (3.5, -4.5) {};
		\node [style=X phase dot bb] (51) at (14.25, -0.25) {$\;e_{13}\pi\;$};
		\node [style=X dot] (63) at (10.75, 1.75) {};
		\node [style=Z dot] (57) at (15.5, -4.5) {};
		\node [style=X dot bb] (12) at (-3.25, -5.5) {};
		\node [style=X dot] (69) at (17.25, 1.75) {};
		\node [style=X dot] (40) at (2.75, 1.75) {};
		\node [style=X dot] (71) at (19.5, 1.75) {};
		\node [style=none] (78) at (18.75, 2.5) {};
		\node [style=X phase dot bb] (53) at (14.25, -4.5) {$\;e_{15}\pi\;$};
		\node [style=X phase dot bb] (23) at (-1, -4.5) {$\;e_3\pi\;$};
		\node [style=Z dot bb] (16) at (-2.25, 0.75) {};
		\node [style=X dot bb] (37) at (10.5, -0.25) {};
		\node [style=X phase dot bb] (44) at (11.75, -3.5) {$\;e_{11}\pi\;$};
		\node [style=X phase dot bb] (42) at (11.75, 0.75) {$\;e_9\pi\;$};
		\node [style=X dot] (68) at (16.5, 1.75) {};
		\node [style=none] (75) at (10.75, 2.5) {};
		\node [style=Z dot] (33) at (2.75, -0.25) {};
		\node [style=Z dot] (48) at (15.5, -3.5) {};
		\node [style=X phase dot bb] (21) at (-1, -0.25) {$\;e_1\pi\;$};
		\node [style=Z dot bb] (26) at (10.5, -3.5) {};
		\node [style=X dot] (41) at (3.5, 1.75) {};
		\node [style=Z dot] (54) at (15.5, -1.25) {};
		\node [style=X dot bb] (47) at (13, -0.25) {};
		\node [style=X dot bb] (39) at (10.5, -4.5) {};
		\node [style=Z dot] (55) at (15.5, -0.25) {};
		\node [style=Z dot bb] (32) at (13, -1.25) {};
		\node [style=none] (72) at (2.75, 2.5) {};
		\node [style=X dot bb] (36) at (9.75, -0.25) {};
		\node [style=X dot] (62) at (10, 1.75) {};
		\node [style=X phase dot bb] (30) at (1.5, -5.5) {$\;e_7\pi\;$};
		\node [style=Z dot bb] (34) at (13, -5.5) {};
		\node [style=Z dot bb] (19) at (0.25, -5.5) {};
		\node [style=Z dot] (56) at (15.5, -5.5) {};
		\node [style=X phase dot bb] (31) at (1.5, -4.5) {$\;e_6\pi\;$};
		\node [style=X dot bb] (8) at (-3.25, -3.5) {};
		\node [style=Z dot bb] (80) at (4, 0.75) {};
		\node [style=X dot bb] (81) at (4, -1.25) {};
		\node [style=X phase dot bb] (82) at (5, 0.75) {$\frac{\pi}{8}$};
		\node [style=Z phase dot bb] (83) at (6.25, 0.75) {$\;e_8\pi\;$};
		\node [style=X phase dot bb] (84) at (7.5, 0.75) {$\frac{\pi}{8}$};
		\node [style=Z dot bb] (85) at (8.5, 0.75) {};
		\node [style=X dot bb] (86) at (8.5, -1.25) {};
		\node [style=Z dot bb] (87) at (9.5, 0.75) {};
		\node [style=Z dot bb] (88) at (9.5, -1.25) {};
		\node [style=X dot bb] (89) at (9.5, -5.5) {};
		\node [style=X dot bb] (90) at (9.5, -3.5) {};
		\node [style=X dot bb] (91) at (-3.25, -0.25) {};
		\node [style=X dot bb] (92) at (-3.25, -1.25) {};
		\node [style=X dot bb] (93) at (-3.25, 0.75) {};
		\node [style=none] (94) at (2.75, 3.25) {$D_0$};
		\node [style=none] (95) at (3.75, 3.25) {$D_1$};
		\node [style=none] (96) at (11, 3.25) {$D_3$};
		\node [style=none] (97) at (10, 3.25) {$D_2$};
		\node [style=none] (98) at (17.5, 3.25) {$D_5$};
		\node [style=none] (99) at (16.5, 3.25) {$D_4$};
		\node [style=none] (100) at (19.75, 3.25) {$O_1$};
		\node [style=none] (101) at (18.75, 3.25) {$O_0$};
	\end{pgfonlayer}
	\begin{pgfonlayer}{edgelayer}
		\draw [style=boldedge] (38) to (39);
		\draw [style=boldedge] (49) to (53);
		\draw [style=boldedge] (52) to (56);
		\draw [bend right=15] (46) to (68);
		\draw [style=boldedge] (22) to (26);
		\draw [style=boldedge] (27) to (31);
		\draw [style=boldedge] (25) to (29);
		\draw [style=boldedge] (10) to (23);
		\draw [style=boldedge] (10) to (17);
		\draw (70) to (78);
		\draw [style=boldedge] (29) to (33);
		\draw [style=boldedge] (17) to (22);
		\draw [style=boldedge] (50) to (54);
		\draw [style=boldedge] (24) to (42);
		\draw [style=boldedge] (24) to (37);
		\draw [style=boldedge] (20) to (24);
		\draw [style=boldedge] (43) to (47);
		\draw [style=boldedge] (18) to (28);
		\draw [out=90, in=270, style=boldedge] (18) to (25);
		\draw [style=boldedge] (45) to (49);
		\draw [style=boldedge] (28) to (32);
		\draw [style=boldedge] (4) to (21);
		\draw [style=boldedge] (4) to (16);
		\draw [style=boldedge] (9) to (10);
		\draw [looseness=0.5, bend right=15] (35) to (63);
		\draw [style=boldedge] (51) to (55);
		\draw (63) to (75);
		\draw [out=150, in=-90, looseness=1.5] (57) to (63);
		\draw [looseness=0.25, bend right=15] (57) to (69);
		\draw [style=boldedge] (12) to (19);
		\draw (69) to (77);
		\draw (40) to (72);
		\draw (71) to (79);
		\draw [style=boldedge] (53) to (57);
		\draw [style=boldedge] (23) to (27);
		\draw [style=boldedge] (16) to (20);
		\draw [style=boldedge] (37) to (43);
		\draw [style=boldedge] (44) to (48);
		\draw [style=boldedge] (42) to (46);
		\draw (68) to (76);
		\draw (33) to (40);
		\draw [out=0, in=-120, looseness=0.75] (33) to (62);
		\draw [looseness=0.25, bend right=15] (48) to (69);
		\draw [style=boldedge] (21) to (25);
		\draw [style=boldedge] (26) to (44);
		\draw [style=boldedge] (26) to (39);
		\draw (41) to (73);
		\draw [looseness=0.5, bend right=15] (54) to (68);
		\draw [bend right=15] (54) to (70);
		\draw [style=boldedge] (47) to (51);
		\draw [style=boldedge] (39) to (45);
		\draw [out=105, in=-30, looseness=0.75] (55) to (62);
		\draw [looseness=0.5, bend right=15] (55) to (68);
		\draw [style=boldedge] (32) to (50);
		\draw [style=boldedge] (32) to (47);
		\draw [style=boldedge] (36) to (37);
		\draw (62) to (74);
		\draw [style=boldedge] (30) to (34);
		\draw [style=boldedge] (34) to (52);
		\draw [style=boldedge] (34) to (49);
		\draw [style=boldedge] (19) to (30);
		\draw [style=boldedge] (19) to (27);
		\draw [looseness=0.25, out=45, in=270] (56) to (69);
		\draw [looseness=0.75, bend right=15] (56) to (71);
		\draw [style=boldedge] (31) to (35);
		\draw [style=boldedge] (8) to (17);
		\draw [style=boldedge] (81) to (80);
		\draw [style=boldedge] (86) to (85);
		\draw [looseness=0.75, bend left, style=boldedge] (89) to (88);
		\draw [looseness=0.75, out=120, in=240, style=boldedge] (90) to (87);
		\draw [style=boldedge] (92) to (18);
		\draw [style=boldedge] (91) to (4);
		\draw [style=boldedge] (93) to (16);
		\draw (41) to (35);
	\end{pgfonlayer}
\end{tikzpicture}
  \label{eqn:rep2}
\end{equation}

A key insight underpinning Tsim's simulation pipeline is that ZX reduction
of the detector sampling diagram yields a
much simpler diagram comprising detector and observable components.
For the example diagram \eqref{eqn:rep2}, we get the following equivalent diagrammatic representation:
\begin{equation}
  \begin{tikzpicture}
	\begin{pgfonlayer}{nodelayer}
		\node [style=X phase dot bb] (1) at (2.25, -0.5) {$\frac{\pi}{8}$};
		\node [style=X phase dot] (9) at (-1, 1) {$\;f_1\pi\;$};
		\node [style=X phase dot] (43) at (6, 1) {$\;f_5\pi\;$};
		\node [style=X phase dot] (49) at (9.5, 1) {$\;f_7\pi\;$};
		\node [style=X phase dot] (59) at (1, 1) {$\;f_2\pi\;$};
		\node [style=X phase dot] (61) at (4.5, 1) {$\;f_4\pi\;$};
		\node [style=X phase dot] (67) at (-2.5, 1) {$\;f_0\pi\;$};
		\node [style=X phase dot] (96) at (8, 1) {$\;f_6\pi\;$};
		\node [style=X phase dot] (102) at (2.5, 1) {$\;f_3\pi\;$};
		\node [style=none] (124) at (-2.5, 2) {};
		\node [style=none] (125) at (-1, 2) {};
		\node [style=none] (126) at (1, 2) {};
		\node [style=none] (127) at (2.5, 2) {};
		\node [style=none] (128) at (4.5, 2) {};
		\node [style=none] (129) at (6, 2) {};
		\node [style=none] (130) at (9.5, 2) {};
		\node [style=none] (131) at (8, 2) {};
		\node [style=X phase dot bb] (132) at (2.25, -1.5) {};
		\node [style=X dot bb] (133) at (6.25, -1.5) {};
		\node [style=Z dot bb] (134) at (6.25, -0.5) {};
		\node [style=Z dot] (135) at (7.5, -0.5) {};
		\node [style=Z dot] (136) at (7.5, -1.5) {};
		\node [style=Z phase dot bb] (137) at (3.75, -0.5) {$\;f_8\pi\;$};
		\node [style=X phase dot bb] (138) at (5.25, -0.5) {$\frac{\pi}{8}$};
		\node [style=none] (94) at (-2.5, 2.75) {$D_0$};
		\node [style=none] (95) at (-1, 2.75) {$D_1$};
		\node [style=none] (139) at (2.5, 2.75) {$D_3$};
		\node [style=none] (97) at (1, 2.75) {$D_2$};
		\node [style=none] (98) at (6, 2.75) {$D_5$};
		\node [style=none] (99) at (4.5, 2.75) {$D_4$};
		\node [style=none] (100) at (9.5, 2.75) {$O_1$};
		\node [style=none] (101) at (8, 2.75) {$O_0$};
	\end{pgfonlayer}
	\begin{pgfonlayer}{edgelayer}
		\draw (9) to (125);
		\draw (43) to (129);
		\draw (49) to (130);
		\draw (59) to (126);
		\draw (61) to (128);
		\draw (67) to (124);
		\draw (96) to (131);
		\draw (102) to (127);
		\draw [style=boldedge] (1) to (134);
		\draw [style=boldedge] (134) to (133);
		\draw [style=boldedge] (132) to (133);
		\draw [style=boldedge] (134) to (135);
		\draw [style=boldedge] (133) to (136);
		\draw [bend right=15] (135) to (96);
		\draw [bend right] (136) to (49);
	\end{pgfonlayer}
\end{tikzpicture}
  \label{eqn:rep_reduced2}
\end{equation}
Here, we have introduced new binary variables $f_i$
that linearly depend on the $e_i$ as
\begin{align}
  f_i = T_{ij} e_j \;\bmod\; 2 \, .
  \label{eqn:detector_sampling}
\end{align}
In general, there are multiple ways to introduce the $f_i$ variables. We choose a minimal set of $f_i$ by performing Gaussian elimination over $\FF$ on the binary matrix of Pauli variable dependencies, retaining only a linearly independent basis.

Importantly, we see that ZX reduction has separated the diagram into two distinct parts: the detectors and the logical observables. The separation is a consequence of the absence
of quantum correlations between the two components.
In many cases, detectors and observables are only classically correlated via
the shared $e_i$ variables in Eq.~\eqref{eqn:detector_sampling}.
This allows us to evaluate the two parts independently and significantly
simplify the sampling process.

\subsubsection{Detectors}
\label{sec:detector-sampling}

The detector part of Eq.~\eqref{eqn:rep_reduced2} factors into 6 individual vertices, each connected only to an output:
\begin{equation}
  \begin{tikzpicture}
	\begin{pgfonlayer}{nodelayer}
		\node [style=X phase dot] (9) at (-1, 0.25) {$\;f_1\pi\;$};
		\node [style=X phase dot] (43) at (6, 0.25) {$\;f_5\pi\;$};
		\node [style=X phase dot] (59) at (1, 0.25) {$\;f_2\pi\;$};
		\node [style=X phase dot] (61) at (4.5, 0.25) {$\;f_4\pi\;$};
		\node [style=X phase dot] (67) at (-2.5, 0.25) {$\;f_0\pi\;$};
		\node [style=X phase dot] (102) at (2.5, 0.25) {$\;f_3\pi\;$};
		\node [style=none] (124) at (-2.5, 1.25) {};
		\node [style=none] (125) at (-1, 1.25) {};
		\node [style=none] (126) at (1, 1.25) {};
		\node [style=none] (127) at (2.5, 1.25) {};
		\node [style=none] (128) at (4.5, 1.25) {};
		\node [style=none] (129) at (6, 1.25) {};
		\node [style=none] (94) at (-2.5, 2) {$D_0$};
		\node [style=none] (95) at (-1, 2) {$D_1$};
		\node [style=none] (139) at (2.5, 2) {$D_3$};
		\node [style=none] (97) at (1, 2) {$D_2$};
		\node [style=none] (98) at (6, 2) {$D_5$};
		\node [style=none] (99) at (4.5, 2) {$D_4$};
	\end{pgfonlayer}
	\begin{pgfonlayer}{edgelayer}
		\draw (9) to (125);
		\draw (43) to (129);
		\draw (59) to (126);
		\draw (61) to (128);
		\draw (67) to (124);
		\draw (102) to (127);
	\end{pgfonlayer}
\end{tikzpicture}
  \label{eqn:dets-only}
\end{equation}
Therefore, each detector bit is directly determined by a single $f_i$.
In turn, the $f_i$ are related to the error bits $e_i$ via the binary matrix $T_{ij}$.
Each column $j$ of $T_{ij}$ corresponds to an error location in the circuit,
and $i$ corresponds to a detector or observable. A given detector $i$ is flipped
by an error $j$ if $T_{ij} = 1$. Note that the matrix $T_{ij}$, obtained here purely through ZX rewriting, is equivalent to the bipartite graph (often referred to as the Tanner graph) of the detector error model constructed by Stim~\cite{gidney2021stim}.

Error bits $e_j$ are sampled from a probability distribution $p(e_{i_k}, \ldots, e_{i_{k+n}})$ of the associated noise channel. A single distribution governs $e_i$ for $n$ columns, where $n$ is typically small (e.g., $n=1$ for a bit-flip channel or $n=4$ for two-qubit depolarizing channels, see Eq.~\eqref{eqn:channel_pauli2}). The detector sampling problem thus reduces to sampling binary variables $e_i$ from their respective error channels
and computing $f_i$ as the $\FF$ matrix multiplication of Eq.~\eqref{eqn:detector_sampling}.

In practice, many error locations may share the same column signature in $T_{ij}$ (i.e.\ the same detector--observable flip pattern). Multiple such error locations can be combined into an \textit{error mechanism}, provided the associated probability distributions are updated accordingly. Tsim performs a series of channel reduction steps that are detailed in Appendix~\ref{sec:channel_reduction}.

Furthermore, following the same strategy as Stim~\cite{gidney2021stim}, Tsim optimizes error sampling for the low-noise regime by exploiting the fact that each error mechanism fires with probability $\perr \ll 1$. Rather than querying each of the $\nerr$ error mechanisms independently for each shot, Tsim samples the gap between successive firing events from a geometric distribution, reducing the expected per-shot cost from $\bigO{\nerr \navgflip}$ to $\bigO{\perr \nerr \navgflip + 1}$, where $\perr$ is the error probability and $\navgflip$ is the average number of detectors flipped by a single error mechanism.

\subsubsection{Observables}

Next, we consider the observable part of the diagram, which consists of a single connected component. The observable diagram represents the logical circuit on
the two logical qubits:
\begin{equation}
  \begin{tikzpicture}
	\begin{pgfonlayer}{nodelayer}
		\node [style=X phase dot bb] (1) at (2.25, -0.5) {$\frac{\pi}{8}$};
		\node [style=X phase dot] (49) at (9, -1.5) {$\;f_7\pi\;$};
		\node [style=X phase dot] (96) at (9, -0.5) {$\;f_6\pi\;$};
		\node [style=none] (130) at (10.5, -1.5) {};
		\node [style=none] (131) at (10.5, -0.5) {};
		\node [style=X phase dot bb] (132) at (2.25, -1.5) {};
		\node [style=X dot bb] (133) at (6.25, -1.5) {};
		\node [style=Z dot bb] (134) at (6.25, -0.5) {};
		\node [style=Z dot] (135) at (7.5, -0.5) {};
		\node [style=Z dot] (136) at (7.5, -1.5) {};
		\node [style=Z phase dot bb] (137) at (3.75, -0.5) {$\;f_8\pi\;$};
		\node [style=X phase dot bb] (138) at (5.25, -0.5) {$\frac{\pi}{8}$};
		\node [style=none] (100) at (11.5, -1.5) {$O_1$};
		\node [style=none] (101) at (11.5, -0.5) {$O_0$};
	\end{pgfonlayer}
	\begin{pgfonlayer}{edgelayer}
		\draw (49) to (130);
		\draw (96) to (131);
		\draw [style=boldedge] (1) to (134);
		\draw [style=boldedge] (134) to (133);
		\draw [style=boldedge] (132) to (133);
		\draw [style=boldedge] (134) to (135);
		\draw [style=boldedge] (133) to (136);
		\draw (135) to (96);
		\draw (136) to (49);
	\end{pgfonlayer}
\end{tikzpicture}
  \label{eqn:obs-only}
\end{equation}
In the Clifford case, for deterministic observables in the absence of noise, this part of the diagram
would also have reduced to individual single-vertex components that could be sampled as described in Sec.~\ref{sec:detector-sampling}.

When there are quantum correlations between observable bits, i.e., when the circuit has
a non-trivial logical action, the observable diagram will be composed of one or more connected components. Here, the sampling strategy proceeds in three steps. First, for each connected component, an autoregressive chain of diagrams is constructed that corresponds to the marginal probabilities of its output bits.
If a connected component has $n$ output bits, one constructs $n$ marginal probability diagrams sequentially (Algorithm~\ref{alg:autoregressive}, \cite{Sutcliffe_2025,kissinger2022simulating}). Second, each marginal diagram is evaluated via stabilizer rank decomposition~\cite{kissinger2022classical,kissinger2022simulating,bravyi2019simulation} (Appendix~\ref{sec:decomp}), which expresses it as a linear combination of parameterized Clifford ZX diagrams. Third, each Clifford diagram in this sum is evaluated as a symbolic expression over the noise parameters $f_i$ \cite{Sutcliffe_2025}, as described in Sec.~\ref{sec:paramzx}.


In the present example, we have a single connected component, and autoregressive sampling is performed as follows:
After isolating the observable component from the rest of the ZX graph, we first discard any accumulated scalar phase terms. These terms appear to depend parametrically on the $e_i$ variables, but they collectively form a normalization factor whose parameter dependence cancels. This normalization factor can be restored by evaluating the diagram with all outputs traced out:
\begin{equation}
  \begin{tikzpicture}
	\begin{pgfonlayer}{nodelayer}
		\node [style=X phase dot bb] (1) at (2.25, -0.5) {$\frac{\pi}{8}$};
		\node [style=X phase dot] (49) at (9, -1.5) {$\;f_7\pi\;$};
		\node [style=X phase dot] (96) at (9, -0.5) {$\;f_6\pi\;$};
		\node [style=Z dot] (130) at (10.75, -1.5) {};
		\node [style=Z dot] (131) at (10.75, -0.5) {};
		\node [style=X phase dot bb] (132) at (2.25, -1.5) {};
		\node [style=X dot bb] (133) at (6.25, -1.5) {};
		\node [style=Z dot bb] (134) at (6.25, -0.5) {};
		\node [style=Z dot] (135) at (7.5, -0.5) {};
		\node [style=Z dot] (136) at (7.5, -1.5) {};
		\node [style=Z phase dot bb] (137) at (3.75, -0.5) {$\;f_8\pi\;$};
		\node [style=X phase dot bb] (138) at (5.25, -0.5) {$\frac{\pi}{8}$};
		\node [style=none] (139) at (0.5, -1) {$P=$};
	\end{pgfonlayer}
	\begin{pgfonlayer}{edgelayer}
		\draw (49) to (130);
		\draw (96) to (131);
		\draw [style=boldedge] (1) to (134);
		\draw [style=boldedge] (134) to (133);
		\draw [style=boldedge] (132) to (133);
		\draw [style=boldedge] (134) to (135);
		\draw [style=boldedge] (133) to (136);
		\draw (135) to (96);
		\draw (136) to (49);
	\end{pgfonlayer}
\end{tikzpicture}
\end{equation}

Next, we compute $P(m_0=0)$, the probability that the first observable is 0, and sample a bit $\tilde{m}_0$ from this distribution.
\begin{equation}
  \begin{tikzpicture}
	\begin{pgfonlayer}{nodelayer}
		\node [style=X phase dot bb] (1) at (2.25, -0.5) {$\frac{\pi}{8}$};
		\node [style=X phase dot] (49) at (9, -1.5) {$\;f_7\pi\;$};
		\node [style=X phase dot] (96) at (9, -0.5) {$\;f_6\pi\;$};
		\node [style=Z dot] (130) at (10.75, -1.5) {};
		\node [style=X dot] (131) at (10.75, -0.5) {};
		\node [style=X phase dot bb] (132) at (2.25, -1.5) {};
		\node [style=X dot bb] (133) at (6.25, -1.5) {};
		\node [style=Z dot bb] (134) at (6.25, -0.5) {};
		\node [style=Z dot] (135) at (7.5, -0.5) {};
		\node [style=Z dot] (136) at (7.5, -1.5) {};
		\node [style=Z phase dot bb] (137) at (3.75, -0.5) {$\;f_8\pi\;$};
		\node [style=X phase dot bb] (138) at (5.25, -0.5) {$\frac{\pi}{8}$};
		\node [style=none] (139) at (-1.5, -1) {$P(m_0=0) = \frac{1}{P}$};
	\end{pgfonlayer}
	\begin{pgfonlayer}{edgelayer}
		\draw (49) to (130);
		\draw (96) to (131);
		\draw [style=boldedge] (1) to (134);
		\draw [style=boldedge] (134) to (133);
		\draw [style=boldedge] (132) to (133);
		\draw [style=boldedge] (134) to (135);
		\draw [style=boldedge] (133) to (136);
		\draw (135) to (96);
		\draw (136) to (49);
	\end{pgfonlayer}
\end{tikzpicture}
\end{equation}
Then, we compute the conditional probability $P(m_1=0 | m_0=\tilde{m}_0) = P(\tilde{m}_0, m_1=0) / P(\tilde{m}_0)$ and sample the observable bit $\tilde{m}_1$:
\begin{equation}
  \begin{tikzpicture}
	\begin{pgfonlayer}{nodelayer}
		\node [style=X phase dot bb] (1) at (2.25, -0.5) {$\frac{\pi}{8}$};
		\node [style=X phase dot] (49) at (9, -1.5) {$\;f_7\pi\;$};
		\node [style=X phase dot] (96) at (9, -0.5) {$\;f_6\pi\;$};
		\node [style=X dot] (130) at (10.75, -1.5) {};
		\node [style=X phase dot] (131) at (10.75, -0.5) {$\;m_0\;$};
		\node [style=X phase dot bb] (132) at (2.25, -1.5) {};
		\node [style=X dot bb] (133) at (6.25, -1.5) {};
		\node [style=Z dot bb] (134) at (6.25, -0.5) {};
		\node [style=Z dot] (135) at (7.5, -0.5) {};
		\node [style=Z dot] (136) at (7.5, -1.5) {};
		\node [style=Z phase dot bb] (137) at (3.75, -0.5) {$\;f_8\pi\;$};
		\node [style=X phase dot bb] (138) at (5.25, -0.5) {$\frac{\pi}{8}$};
		\node [style=none] (139) at (-2, -1) {$P(m_0, m_1=0) = \frac{1}{P}$};
	\end{pgfonlayer}
	\begin{pgfonlayer}{edgelayer}
		\draw (49) to (130);
		\draw (96) to (131);
		\draw [style=boldedge] (1) to (134);
		\draw [style=boldedge] (134) to (133);
		\draw [style=boldedge] (132) to (133);
		\draw [style=boldedge] (134) to (135);
		\draw [style=boldedge] (133) to (136);
		\draw (135) to (96);
		\draw (136) to (49);
	\end{pgfonlayer}
\end{tikzpicture}
  \label{eqn:rep_reduced2_marginal_2}
\end{equation}
Each marginal probability diagram is itself a parameterized ZX diagram that is evaluated via stabilizer rank decomposition as a linear combination of Clifford terms. For $n$ observables, $n$ such diagrams plus an additional normalization factor need to be evaluated. As in Eq.~\eqref{eqn:rep_reduced2_marginal_2}, diagrams depend on previously sampled bits. The dependence on these bits is retained as parameterized vertices in the diagram~\cite{Sutcliffe_2025,sutcliffe2025novel}, allowing the same diagram to be reused across bit configurations and enabling efficient batched sampling of many shots in parallel (Sec.~\ref{sec:paramzx}).

\subsubsection{The advantage of sampling detectors over measurements}
\label{sec:detector_advantage}

Weak simulation of quantum circuits, i.e.\ sampling bit strings from ZX diagrams, requires autoregressive evaluation of marginal probabilities (Algorithm~\ref{alg:autoregressive}, \cite{kissinger2022simulating}). When sampling raw measurement outcomes, this introduces an overhead proportional to the number of measurements $N_m$, since each measurement bit requires evaluating an additional diagram.

Detector sampling largely avoids this overhead. In a well-designed fault-tolerant protocol, detectors are deterministic in the absence of noise, and ZX reduction maps each detector to an individual connected component consisting of a single internal vertex. Pauli noise channels may flip detectors and introduce correlations, but these correlations are captured classically through correlated sampling of the $f_i$ phase parameters and do not add edges to the ZX graph. Consequently, detector bits are obtained directly from the sampled $f_i$ configuration via Eq.~\eqref{eqn:detector_sampling}, with no autoregressive overhead.
The autoregressive procedure is then needed only for the observable bits. However, in typical QEC experiments, there are $\bigO{1}$ observables.

If we had sampled measurements, instead of detectors and observables, the ZX diagram would generally be a single connected component, reflecting the correlations of measurement bits. Compared to this case, detector sampling effectively reduces the per-shot sampling cost by a factor of $N_m$.

We note that this separation is a heuristic property: ZX reduction may not cleanly split all detectors from the observables in the presence of non-Clifford gates, in which case some detectors may have to be evaluated autoregressively.

\subsubsection{Complexity}

Under the assumption that detectors and observables separate after ZX reduction and that the number of observables is $\bigO{1}$, we summarize the asymptotic cost of simulation. Let $\ngates$ denote the number of Clifford gates in the circuit, $\nmagic$ the number of non-Clifford gates, $\nerr$ the number of error mechanisms (typically $\bigO{\ngates}$), $\navgflip$ the average number of detectors flipped per mechanism (typically $\bigO{1}$), $\perr$ the physical error probability, $\batchsize$ the shot batch size, and $\stabrank = 2^{\decomprate \nmagic}$ the stabilizer rank of the decomposition, where $\decomprate \le 1$ is a constant that depends on the decomposition strategy (Appendix~\ref{sec:decomp}).

\paragraph{Compilation.} The dominant costs are ZX graph reduction, which scales as $\bigO{\ngates^3}$ via local complementation and pivoting, and the assembly of the stabilizer rank decomposition tensors, which scales as $\bigO{\stabrank \nmagic^2}$. The total compilation cost is therefore $\bigO{\ngates^3 + \stabrank \nmagic^2}$. Compilation is performed once and amortized over all subsequent sampling calls.

\paragraph{Sampling.} Each shot requires (i)~sampling error bits and computing detector flips via sparse $\FF$ matrix--vector multiplication at cost $\bigO{\perr \nerr \navgflip}$ (Sec.~\ref{sec:detector-sampling}), and (ii)~evaluating the observable diagram, which involves contracting binary matrices of dimension $\nmagic \times \nmagic \times \stabrank$ with the $\batchsize$-shot parameter batch, at cost $\bigO{\stabrank \nmagic^2 \batchsize}$ per marginal. The per-batch sampling cost is therefore $\bigO{\perr \nerr \navgflip \batchsize + \stabrank \nmagic^2 \batchsize}$. In the low-magic limit of small $\nmagic$, the dominant factor will be a linear dependence on the number of error mechanisms $\nerr$.

\section{Benchmarks}
\label{sec:benchmarks}
\begin{figure}[t]
  \centering
  \includegraphics[width=\linewidth]{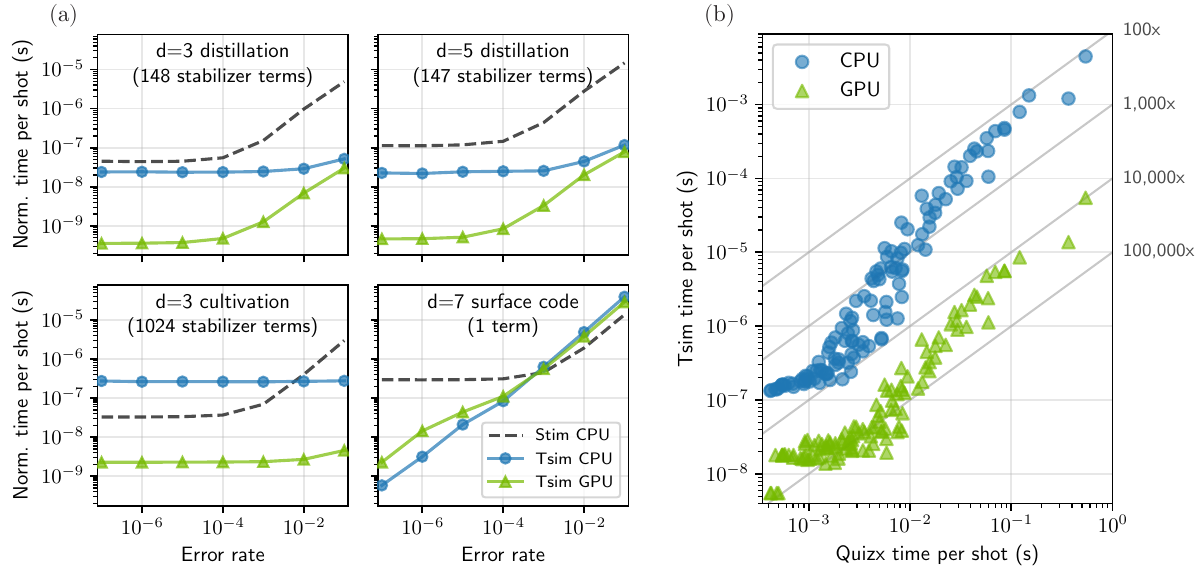}
  \caption{(a) Shot sampling time comparison between Tsim and Stim for
    $d=3,5$ distillation circuits from \cite{sales2025experimental}, $d=3$ cultivation circuit from \cite{gidney2024magic}, and a $d=7$ rotated surface code (7 rounds). Tsim runtime is normalized by the number of stabilizer terms in the decomposition. For reference, unnormalized runtimes are shown in Fig.~\ref{fig:benchmark-unnormalized}. (b) Runtime comparison between Tsim and quizx. Each data point corresponds to a randomly generated circuit of exponentiated Paulis of weight 2--4 and depth between 1 and 21. For Tsim, batch size is autotuned to maximize throughput. All CPU benchmarks were performed on an Apple M4 Pro. GPU benchmarks were conducted on an NVIDIA Grace Hopper GH200 system with a 72-core ARM Neoverse-V2 CPU.
  }
  \label{fig:benchmark}
\end{figure}

To evaluate performance, we performed a series of simulations on an
Apple M4 Pro CPU and an NVIDIA Grace Hopper GH200 system with a 72-core
ARM Neoverse-V2 CPU. For every simulation, we compiled a detector sampler and
autotuned the batch size for maximum throughput. We then recorded the sampling
time after a warmup step that includes the kernel's JAX/XLA just-in-time
compilation.

Fig.~\ref{fig:benchmark}(a) shows Stim (black dashed) and Tsim sampling time on CPU (solid blue) and GPU (solid green) as a function of physical error rate for the
$d=3,5$ distillation circuits of \cite{sales2025experimental}, the $d=3$
cultivation circuit of \cite{gidney2024magic}, and a $d=7$ rotated surface-code
memory with seven rounds. Here, Tsim runtime is normalized by the number of
stabilizer terms in the decomposition. This removes the exponential dependence
on stabilizer rank. For reference, unnormalized data are shown in
Fig.~\ref{fig:benchmark-unnormalized}.

The first three panels show non-Clifford circuits. Since Stim only supports
Clifford operations, we simulated proxy circuits in which T gates are replaced
by S gates and rotations are omitted. Stim shows a clear runtime dependence on
the error rate for $p>10^{-3}$. In this regime, entropy generation for Pauli
channels becomes the bottleneck. Tsim also shows a $p$ dependence, but it is
less pronounced. On CPU, the runtime is dominated by tensor contraction during
diagrammatic evaluation. On GPU, this part is accelerated by roughly two orders
of magnitude, which reveals the $p$ dependence when the stabilizer rank is not
too large.

For purely Clifford diagrams with deterministic detectors, Tsim uses a sparse geometric sampler based on Eq.~\eqref{eqn:detector_sampling}. This path is not GPU-accelerated, so CPU and GPU runtimes are similar, as shown in the bottom right panel of Fig.~\ref{fig:benchmark}(a). It also achieves the asymptotically optimal
$\bigO{\perr}$ scaling, outperforming Stim for $\perr < 10^{-3}$.

Figure~\ref{fig:benchmark}(b) compares Tsim with quizx on random circuits of
exponentiated Pauli gates, following \cite{kissinger2022classical}. We sampled
circuits of 20 qubits with Pauli strings of weight 2--4 and depth 1--21. For both Tsim and quizx, the Cat5 stabilizer rank decomposition (Eq.~\ref{eqn:cat5}) is used.
For Tsim, we additionally inserted a depolarizing channel with $p=0.001$ after
every gate. For each random circuit, we recorded the quizx and Tsim runtime. At
low T count, Tsim on GPU is more than five orders of magnitude faster than
quizx. This speedup reflects the fact that Tsim pays the $\bigO{\ngates^3}$ ZX
simplification cost during compilation rather than during sampling. As the T
count increases, this advantage narrows, but Tsim still achieves speedups of
about $10^2$ on CPU and $10^4$ on GPU.



\section{Discussion}
\label{sec:summary}

We have presented Tsim, a GPU-accelerated simulator for universal noisy quantum circuits targeting quantum error correction, based on parameterized ZX-calculus reduction and stabilizer rank decomposition. Tsim implements the Stim API for seamless integration with existing QEC simulation workflows, and further accepts circuits in other formats, including Bloqade, OpenQASM, Cirq, and QIR. The compile-once, sample-many architecture enables per-stabilizer-term sampling performance on CPU that is comparable to or faster than Stim on both Clifford and non-Clifford practical QEC gadgets. GPU acceleration can further improve throughput by up to two orders of magnitude. Compared to quizx, Tsim achieves speedups of up to five orders of magnitude by amortizing the ZX simplification cost at compile time.

There are several potential directions to be explored in the future.
The primary computational bottleneck of the current approach is the exponential scaling of stabilizer rank with the number of non-Clifford gates. Possible approaches to reduce this cost include cutting decompositions that exploit circuit structure~\cite{wan2025cutting,wan2026simulatingmagicstatecultivation}, more efficient rewrite rules for the doubled diagram representation, and approximate decompositions that trade small errors in the output distribution for significantly reduced rank~\cite{bravyi2019simulation}. On the engineering side, replacing the Python-based ZX reduction with a Rust backend such as quizx~\cite{kissinger2022classical} would accelerate the compilation step, while techniques such as branch merging and extended autoregressive sampling could further improve sampling throughput. Looking further ahead, alternative graphical calculi such as the ZH calculus may enable more efficient native decompositions for non-Clifford gates beyond single-qubit rotations, such as the Toffoli gate~\cite{backens2019zhcomplete}. Finally, extending the noise model beyond Pauli channels to coherent errors, atom loss, and leakage would broaden Tsim's applicability to a wider range of hardware platforms and error models.

\section*{Acknowledgments}


We thank Matthew Sutcliffe, John van de Wetering, Sara Meng Li, and Lara Madison for insightful discussions and Matthew Sutcliffe for making the ParamZX extension~\cite{paramzx2024} to PyZX publicly available, on which parts of Tsim's parameterized ZX reduction pipeline are built. Following the initial release of Tsim, Kwok Ho Wan and Zhenghao Zhong~\cite{wan2026simulatingmagicstatecultivation} developed several extensions to the framework, showcasing significantly improved sampling throughput for magic state cultivation circuits. We added new features based on their optimized cutting decomposition and a fast geometric sampling routine, and plan to integrate
additional components. R.H. would like to thank Phillip Weinberg, Casey Duckering, and Jonathan Wurtz  for fruitful discussions.

\bibliographystyle{unsrt}
\bibliography{references}

\onecolumn\newpage
\appendix

\section{ZX-Calculus}
\subsection{Rewrite rules}
\label{sec:zx}

The ZX-calculus is a graphical tensor network representation for quantum circuits. It represents quantum processes as graphs composed of Z (green) and X (red) vertices, often referred to as \emph{spiders}. Each spider carries a phase parameter $\alpha \in \RR$ and corresponds to a linear map defined as follows:
\begin{equation}
  \begin{tikzpicture}
	\begin{pgfonlayer}{nodelayer}
		\node [style=none] (0) at (-3, 3) {};
		\node [style=none] (1) at (-3, 1.25) {};
		\node [style=Z phase dot] (4) at (-1.5, 2.25) {$\alpha$};
		\node [style=none] (6) at (0, 3) {};
		\node [style=none] (8) at (-0.25, 2.25) {$\vdots$};
		\node [style=none] (9) at (-2.75, 2.25) {$\vdots$};
		\node [style=none] (12) at (-3, 3.5) {};
		\node [style=none] (13) at (0, 3.5) {};
		\node [style=none] (14) at (0, 1.25) {};
		\node [style=none] (16) at (7.5, 2.25) {$=\;|00\ldots 0\rangle \langle 00\ldots 0| +e^{i\alpha}|11\ldots 1\rangle \langle 11\ldots 1|$};
		\node [style=none] (17) at (-3, -0.5) {};
		\node [style=none] (18) at (-3, -2.25) {};
		\node [style=X phase dot] (19) at (-1.5, -1.25) {$\alpha$};
		\node [style=none] (20) at (0, -0.5) {};
		\node [style=none] (21) at (-0.25, -1.25) {$\vdots$};
		\node [style=none] (22) at (-2.75, -1.25) {$\vdots$};
		\node [style=none] (23) at (-3, 0) {};
		\node [style=none] (24) at (0, 0) {};
		\node [style=none] (25) at (0, -2.25) {};
		\node [style=none] (26) at (9.5, -1.25) {$=\;|++\ldots +\rangle \langle ++\ldots +| +e^{i\alpha}|--\ldots -\rangle \langle --\ldots -|$};
	\end{pgfonlayer}
	\begin{pgfonlayer}{edgelayer}
		\draw [bend left] (0.center) to (4);
		\draw [bend left] (4) to (6.center);
		\draw [bend left] (4) to (1.center);
		\draw [bend left, looseness=1.25] (12.center) to (4);
		\draw [bend left, looseness=1.25] (4) to (13.center);
		\draw [bend right=15] (4) to (14.center);
		\draw [bend left] (17.center) to (19);
		\draw [bend left] (19) to (20.center);
		\draw [bend left] (19) to (18.center);
		\draw [bend left, looseness=1.25] (23.center) to (19);
		\draw [bend left, looseness=1.25] (19) to (24.center);
		\draw [bend right=15] (19) to (25.center);
	\end{pgfonlayer}
\end{tikzpicture}
  \label{eqn:tensor_def}
\end{equation}
Edges in ZX diagrams denote tensor contraction (composition of linear maps). Diagram juxtaposition corresponds to the tensor product.
In addition to standard edges, the Hadamard operation is represented as a blue edge.
\begin{equation}
  \begin{tikzpicture}
	\begin{pgfonlayer}{nodelayer}
		\node [style=none] (286) at (0, 0) {$e^{-i\frac{\pi}{4}}$};
		\node [style=Z phase dot] (288) at (2.25, 0) {$\frac{\pi}{2}$};
		\node [style=Z phase dot] (289) at (4.75, 0) {$\frac{\pi}{2}$};
		\node [style=X phase dot] (290) at (3.5, 0) {$\frac{\pi}{2}$};
		\node [style=none] (291) at (5.75, 0) {};
		\node [style=none] (292) at (7, 0) {$\equiv$};
		\node [style=none] (293) at (8.25, 0) {};
		\node [style=none] (294) at (10.75, 0) {};
		\node [style=none] (282) at (13.25, 0) {};
		\node [style=none] (283) at (15.75, 0) {};
		\node [style=hadamard] (284) at (14.5, 0) {};
		\node [style=none] (285) at (12, 0) {$\equiv$};
		\node [style=none] (295) at (1.25, 0) {};
	\end{pgfonlayer}
	\begin{pgfonlayer}{edgelayer}
		\draw (288) to (290);
		\draw (290) to (289);
		\draw (289) to (291.center);
		\draw [style=hadamard edge] (293.center) to (294.center);
		\draw (282.center) to (284);
		\draw (284) to (283.center);
		\draw (295.center) to (288);
	\end{pgfonlayer}
\end{tikzpicture}
  \label{eqn:hadamard}
\end{equation}
The ZX-calculus comes equipped with rewrite rules that preserve the semantics of the represented linear map, allowing diagrams to be transformed algebraically.
Note that we use Latin letters to denote binary variables, whereas Greek letters represent continuous phases.
\begin{equation*}
  \input{./figures/param-rules2.tikz}
  \label{eqn:param_rules}
\end{equation*}

In this work, we are interested in applying ZX rewrite rules to diagrams with parameterized Pauli vertices, i.e., vertices with either phase $0$ or $\pi$. In the Clifford segment, Pauli vertices assume a particularly trivial form: Paulis map to Paulis under Clifford conjugation. As such, Paulis can only flip measurement
outcomes.

An interesting consequence of this
is that Clifford ZX reduction rules are agnostic to the presence of Pauli vertices. Explicitly, the ZX rules can be applied to vertices
with phases $a\pi$ where $a\in[0,1]$.

However, this does not hold for the
$\pi$-commutation rule $(d)$ when it operates on non-Clifford vertices. The rewrite $(d)$ transforms phases $\alpha \mapsto -\alpha = \alpha - 2\alpha$, effectively generating a new term $-2\alpha$. For Clifford circuits, $\alpha$ is a multiple of $\pi/2$, so that $2\alpha$ is indeed a Pauli phase. However, when non-Clifford gates are present, the $\pi$-commutation rule generates phases that are no longer multiples of $\pi$. In diagrammatic notation, we have
\begin{equation*}
  \begin{tikzpicture}
	\begin{pgfonlayer}{nodelayer}
		\node [style=none] (157) at (0, -2) {$=$};
		\node [style=Z phase dot] (158) at (5.5, -2) {$\pi$};
		\node [style=X phase dot] (159) at (7.75, 0) {$a\pi$};
		\node [style=none] (160) at (8.5, -1.75) {$\vdots$};
		\node [style=X phase dot] (161) at (7.75, -1) {$a\pi$};
		\node [style=X phase dot] (162) at (7.75, -3) {$a\pi$};
		\node [style=none] (163) at (4.25, -2) {};
		\node [style=none] (164) at (8.75, 0) {};
		\node [style=none] (165) at (8.75, -1) {};
		\node [style=none] (166) at (8.75, -3) {};
		\node [style=none] (167) at (2.5, -2) {$e^{ia\pi}$};
		\node [style=Z phase dot] (168) at (-3.5, -2) {$\pi$};
		\node [style=none] (169) at (-2, 0) {};
		\node [style=none] (170) at (-1.5, -1.75) {$\vdots$};
		\node [style=none] (171) at (-5.75, -2) {};
		\node [style=none] (172) at (-2, -1) {};
		\node [style=none] (173) at (-2, -3) {};
		\node [style=X phase dot] (174) at (-4.75, -2) {$a\pi$};
		\node [style=none] (175) at (-1.25, 0) {};
		\node [style=none] (176) at (-1.25, -1) {};
		\node [style=none] (177) at (-1.25, -3) {};
		\node [style=none] (181) at (0, -7) {$=$};
		\node [style=Z phase dot] (182) at (5.5, -7) {$\;\frac{\pi}{2}+a\pi\;$};
		\node [style=X phase dot] (183) at (7.75, -5) {$a\pi$};
		\node [style=none] (184) at (8.5, -6.75) {$\vdots$};
		\node [style=X phase dot] (185) at (7.75, -6) {$a\pi$};
		\node [style=X phase dot] (186) at (7.75, -8) {$a\pi$};
		\node [style=none] (187) at (3.5, -7) {};
		\node [style=none] (188) at (8.75, -5) {};
		\node [style=none] (189) at (8.75, -6) {};
		\node [style=none] (190) at (8.75, -8) {};
		\node [style=none] (191) at (2.5, -7) {$e^{ia\frac{\pi}{2}}$};
		\node [style=Z phase dot] (192) at (-3.5, -7) {$\;\frac{\pi}{2}\;$};
		\node [style=none] (193) at (-2, -5) {};
		\node [style=none] (194) at (-1.5, -6.75) {$\vdots$};
		\node [style=none] (195) at (-5.75, -7) {};
		\node [style=none] (196) at (-2, -6) {};
		\node [style=none] (197) at (-2, -8) {};
		\node [style=X phase dot] (198) at (-4.75, -7) {$a\pi$};
		\node [style=none] (199) at (-1.25, -5) {};
		\node [style=none] (200) at (-1.25, -6) {};
		\node [style=none] (201) at (-1.25, -8) {};
		\node [style=none] (202) at (0, -12) {$=$};
		\node [style=Z phase dot] (203) at (5.5, -12) {$\;\frac{\pi}{4}-a\frac{\pi}{2}\;$};
		\node [style=X phase dot] (204) at (7.75, -10) {$a\pi$};
		\node [style=none] (205) at (8.5, -11.75) {$\vdots$};
		\node [style=X phase dot] (206) at (7.75, -11) {$a\pi$};
		\node [style=X phase dot] (207) at (7.75, -13) {$a\pi$};
		\node [style=none] (208) at (3.5, -12) {};
		\node [style=none] (209) at (8.75, -10) {};
		\node [style=none] (210) at (8.75, -11) {};
		\node [style=none] (211) at (8.75, -13) {};
		\node [style=none] (212) at (2.5, -12) {$e^{ia\frac{\pi}{4}}$};
		\node [style=Z phase dot] (213) at (-3.5, -12) {$\;\frac{\pi}{4}\;$};
		\node [style=none] (214) at (-2, -10) {};
		\node [style=none] (215) at (-1.5, -11.75) {$\vdots$};
		\node [style=none] (216) at (-5.75, -12) {};
		\node [style=none] (217) at (-2, -11) {};
		\node [style=none] (218) at (-2, -13) {};
		\node [style=X phase dot] (219) at (-4.75, -12) {$a\pi$};
		\node [style=none] (220) at (-1.25, -10) {};
		\node [style=none] (221) at (-1.25, -11) {};
		\node [style=none] (222) at (-1.25, -13) {};
		\node [style=none] (223) at (-7, -2) {$(d.1)$};
		\node [style=none] (224) at (-7, -7) {$(d.2)$};
		\node [style=none] (225) at (-7, -12) {$(d.3)$};
	\end{pgfonlayer}
	\begin{pgfonlayer}{edgelayer}
		\draw (158) to (163.center);
		\draw (162) to (166.center);
		\draw (161) to (165.center);
		\draw (159) to (164.center);
		\draw (168) to (174);
		\draw (174) to (171.center);
		\draw (173.center) to (177.center);
		\draw (172.center) to (176.center);
		\draw (169.center) to (175.center);
		\draw [in=180, out=-75, looseness=1.25] (168) to (173.center);
		\draw [in=-180, out=75, looseness=1.25] (168) to (172.center);
		\draw [in=-180, out=90, looseness=1.25] (168) to (169.center);
		\draw [in=180, out=90, looseness=1.25] (158) to (159);
		\draw [in=-180, out=75, looseness=1.25] (158) to (161);
		\draw [in=180, out=-75, looseness=1.25] (158) to (162);
		\draw (182) to (187.center);
		\draw (186) to (190.center);
		\draw (185) to (189.center);
		\draw (183) to (188.center);
		\draw (192) to (198);
		\draw (198) to (195.center);
		\draw (197.center) to (201.center);
		\draw (196.center) to (200.center);
		\draw (193.center) to (199.center);
		\draw [in=180, out=-75, looseness=1.25] (192) to (197.center);
		\draw [in=-180, out=75, looseness=1.25] (192) to (196.center);
		\draw [in=-180, out=90, looseness=1.25] (192) to (193.center);
		\draw [in=180, out=90, looseness=1.25] (182) to (183);
		\draw [in=-180, out=75, looseness=1.25] (182) to (185);
		\draw [in=180, out=-75, looseness=1.25] (182) to (186);
		\draw (203) to (208.center);
		\draw (207) to (211.center);
		\draw (206) to (210.center);
		\draw (204) to (209.center);
		\draw (213) to (219);
		\draw (219) to (216.center);
		\draw (218.center) to (222.center);
		\draw (217.center) to (221.center);
		\draw (214.center) to (220.center);
		\draw [in=180, out=-75, looseness=1.25] (213) to (218.center);
		\draw [in=-180, out=75, looseness=1.25] (213) to (217.center);
		\draw [in=-180, out=90, looseness=1.25] (213) to (214.center);
		\draw [in=180, out=90, looseness=1.25] (203) to (204);
		\draw [in=-180, out=75, looseness=1.25] (203) to (206);
		\draw [in=180, out=-75, looseness=1.25] (203) to (207);
	\end{pgfonlayer}
\end{tikzpicture}
  \label{eqn:pi_copy}
\end{equation*}
where $(d.3)$ generates a parameterized Clifford phase $a\pi/2$, which is not a Pauli phase. In practice, we simply ignore the $\pi$-commutation rule when non-Clifford vertices are involved.

The above rules can be used to derive two additional rules, referred to as the \emph{local complementation} rule
\begin{equation}
  \begin{tikzpicture}
	\begin{pgfonlayer}{nodelayer}
		\node [style=Z phase dot] (0) at (-4.5, 0) {$\;\frac{\pi}{2}+b\pi\;$};
		\node [style=none] (1) at (-2.75, 0.25) {$\vdots$};
		\node [style=none] (2) at (-0.75, 0) {$=$};
		\node [style=Z phase dot] (3) at (-6.25, -1) {$\;\alpha_{2}\;$};
		\node [style=Z phase dot] (4) at (-2.75, -1) {$\;\alpha_{3}\;$};
		\node [style=Z phase dot] (5) at (-6.25, 1) {$\;\alpha_{1}\;$};
		\node [style=Z phase dot] (6) at (-2.75, 1) {$\;\alpha_{n}\;$};
		\node [style=none] (7) at (-3.25, 2) {};
		\node [style=none] (8) at (-2.25, 2) {};
		\node [style=none] (9) at (-2.75, 2) {$...$};
		\node [style=none] (10) at (-6.75, 2) {};
		\node [style=none] (11) at (-5.75, 2) {};
		\node [style=none] (12) at (-6.25, 2) {$...$};
		\node [style=none] (13) at (-3.25, -2) {};
		\node [style=none] (14) at (-2.25, -2) {};
		\node [style=none] (15) at (-2.75, -2) {$...$};
		\node [style=none] (16) at (-6.75, -2) {};
		\node [style=none] (17) at (-5.75, -2) {};
		\node [style=none] (18) at (-6.25, -2) {$...$};
		\node [style=none] (19) at (13.25, 0.25) {$\vdots$};
		\node [style=Z phase dot] (20) at (9, -1) {$\;\alpha_2 - \frac{\pi}{2}-b\pi\;$};
		\node [style=Z phase dot] (21) at (13.5, -1) {$\;\alpha_3 - \frac{\pi}{2}-b\pi\;$};
		\node [style=Z phase dot] (22) at (9, 1) {$\;\alpha_1 - \frac{\pi}{2}-b\pi\;$};
		\node [style=Z phase dot] (23) at (13.5, 1) {$\;\alpha_n - \frac{\pi}{2}-b\pi\;$};
		\node [style=none] (24) at (13, 2) {};
		\node [style=none] (25) at (14, 2) {};
		\node [style=none] (26) at (13.5, 2) {$...$};
		\node [style=none] (27) at (8.5, 2) {};
		\node [style=none] (28) at (9.5, 2) {};
		\node [style=none] (29) at (9, 2) {$...$};
		\node [style=none] (30) at (13, -2) {};
		\node [style=none] (31) at (14, -2) {};
		\node [style=none] (32) at (13.5, -2) {$...$};
		\node [style=none] (33) at (8.5, -2) {};
		\node [style=none] (34) at (9.5, -2) {};
		\node [style=none] (35) at (9, -2) {$...$};
		\node [style=none] (36) at (4, 0) {$e^{- ib\frac{\pi}{2}} e^{{i\frac{\pi}{4}}} \sqrt{2}^{\frac{(n-1)(n-2)}{2}}$};
	\end{pgfonlayer}
	\begin{pgfonlayer}{edgelayer}
		\draw [style=hadamard edge] (0) to (5);
		\draw [style=hadamard edge] (0) to (6);
		\draw [style=hadamard edge] (0) to (4);
		\draw [style=hadamard edge] (0) to (3);
		\draw (7.center) to (6);
		\draw (8.center) to (6);
		\draw (5) to (10.center);
		\draw (5) to (11.center);
		\draw (3) to (16.center);
		\draw (3) to (17.center);
		\draw (4) to (13.center);
		\draw (4) to (14.center);
		\draw (24.center) to (23);
		\draw (25.center) to (23);
		\draw (22) to (27.center);
		\draw (22) to (28.center);
		\draw (20) to (33.center);
		\draw (20) to (34.center);
		\draw (21) to (30.center);
		\draw (21) to (31.center);
		\draw [style=hadamard edge] (22) to (20);
		\draw [style=hadamard edge] (20) to (21);
		\draw [style=hadamard edge] (21) to (23);
		\draw [style=hadamard edge] (23) to (22);
		\draw [style=hadamard edge] (22) to (21);
		\draw [style=hadamard edge] (23) to (20);
	\end{pgfonlayer}
\end{tikzpicture}
  \label{eqn:local_comp}
\end{equation}
and the \emph{pivoting} rule
\begin{equation}
  \begin{tikzpicture}
	\begin{pgfonlayer}{nodelayer}
		\node [style=Z phase dot] (37) at (-0.75, -2) {$b\pi$};
		\node [style=none] (38) at (0.5, -1.75) {$\vdots$};
		\node [style=none] (39) at (2, -2.5) {$=$};
		\node [style=Z phase dot] (40) at (-4.75, -1) {$\;\alpha_{1}\;$};
		\node [style=none] (41) at (-5.25, 0) {};
		\node [style=none] (42) at (-4.25, 0) {};
		\node [style=none] (43) at (-4.75, 0) {$...$};
		\node [style=Z phase dot] (44) at (-3.25, -2) {$a\pi$};
		\node [style=Z phase dot] (45) at (-4.75, -3) {$\;\alpha_{n}\;$};
		\node [style=none] (46) at (-5.25, -4) {};
		\node [style=none] (47) at (-4.25, -4) {};
		\node [style=none] (48) at (-4.75, -4) {$...$};
		\node [style=Z phase dot] (49) at (0.5, -1) {$\;\gamma_{1}\;$};
		\node [style=none] (50) at (0, 0) {};
		\node [style=none] (51) at (1, 0) {};
		\node [style=none] (52) at (0.5, 0) {$...$};
		\node [style=Z phase dot] (53) at (0.5, -3) {$\;\gamma_{l}\;$};
		\node [style=none] (54) at (0, -4) {};
		\node [style=none] (55) at (1, -4) {};
		\node [style=none] (56) at (0.5, -4) {$...$};
		\node [style=Z phase dot] (57) at (-2, -3.25) {$\;\beta_{1}\;$};
		\node [style=none] (58) at (-2.75, -4) {};
		\node [style=none] (59) at (-1.25, -4) {};
		\node [style=none] (60) at (-2, -4) {$...$};
		\node [style=Z phase dot] (61) at (-2, -5.5) {$\;\beta_{m}\;$};
		\node [style=none] (62) at (-2.75, -6.25) {};
		\node [style=none] (63) at (-1.25, -6.25) {};
		\node [style=none] (64) at (-2, -6.25) {$...$};
		\node [style=none] (65) at (-4.75, -1.75) {$\vdots$};
		\node [style=none] (66) at (-2, -4.25) {$\vdots$};
		\node [style=none] (67) at (4.75, -2.5) {$e^{iab\pi}\sqrt{2}^{E}$};
		\node [style=none] (68) at (14.25, -1.75) {$\vdots$};
		\node [style=Z phase dot] (69) at (8.25, -1) {$\;\alpha_{1}+b\pi\;$};
		\node [style=none] (70) at (7.75, 0) {};
		\node [style=none] (71) at (8.75, 0) {};
		\node [style=none] (72) at (8.25, 0) {$...$};
		\node [style=Z phase dot] (73) at (8.25, -3) {$\;\alpha_{n}+b\pi\;$};
		\node [style=none] (74) at (7.75, -4) {};
		\node [style=none] (75) at (8.75, -4) {};
		\node [style=none] (76) at (8.25, -4) {$...$};
		\node [style=Z phase dot] (77) at (14.25, -1) {$\;\gamma_{1}+a\pi\;$};
		\node [style=none] (78) at (13.75, 0) {};
		\node [style=none] (79) at (14.75, 0) {};
		\node [style=none] (80) at (14.25, 0) {$...$};
		\node [style=Z phase dot] (81) at (14.25, -3) {$\;\gamma_{l}+a\pi\;$};
		\node [style=none] (82) at (13.75, -4) {};
		\node [style=none] (83) at (14.75, -4) {};
		\node [style=none] (84) at (14.25, -4) {$...$};
		\node [style=Z phase dot] (85) at (11.25, -4.75) {$\;\beta_{1}+a\pi+b\pi+\pi\;$};
		\node [style=none] (86) at (10.5, -5.5) {};
		\node [style=none] (87) at (12, -5.5) {};
		\node [style=none] (88) at (11.25, -5.5) {$...$};
		\node [style=Z phase dot] (89) at (11.25, -7) {$\;\beta_{m}+a\pi+b\pi+\pi\;$};
		\node [style=none] (90) at (10.5, -7.75) {};
		\node [style=none] (91) at (12, -7.75) {};
		\node [style=none] (92) at (11.25, -7.75) {$...$};
		\node [style=none] (93) at (8.25, -1.75) {$\vdots$};
		\node [style=none] (94) at (11.25, -5.75) {$\vdots$};
	\end{pgfonlayer}
	\begin{pgfonlayer}{edgelayer}
		\draw (40) to (41.center);
		\draw (40) to (42.center);
		\draw (45) to (46.center);
		\draw (45) to (47.center);
		\draw (49) to (50.center);
		\draw (49) to (51.center);
		\draw (53) to (54.center);
		\draw (53) to (55.center);
		\draw (57) to (58.center);
		\draw (57) to (59.center);
		\draw (61) to (62.center);
		\draw (61) to (63.center);
		\draw [style=hadamard edge] (44) to (40);
		\draw [style=hadamard edge] (44) to (45);
		\draw [style=hadamard edge] (44) to (37);
		\draw [style=hadamard edge] (37) to (49);
		\draw [style=hadamard edge] (37) to (53);
		\draw [style=hadamard edge, bend right] (44) to (57);
		\draw [style=hadamard edge, bend right] (57) to (37);
		\draw [style=hadamard edge, in=165, out=-105, looseness=0.75] (44) to (61);
		\draw [style=hadamard edge, in=285, out=15, looseness=0.75] (61) to (37);
		\draw (69) to (70.center);
		\draw (69) to (71.center);
		\draw (73) to (74.center);
		\draw (73) to (75.center);
		\draw [in=300, out=120] (77) to (78.center);
		\draw (77) to (79.center);
		\draw (81) to (82.center);
		\draw (81) to (83.center);
		\draw (85) to (86.center);
		\draw (85) to (87.center);
		\draw (89) to (90.center);
		\draw (89) to (91.center);
		\draw [style=hadamard edge] (69) to (77);
		\draw [style=hadamard edge] (69) to (81);
		\draw [style=hadamard edge] (77) to (73);
		\draw [style=hadamard edge] (69) to (85);
		\draw [style=hadamard edge] (69) to (89);
		\draw [style=hadamard edge] (77) to (85);
		\draw [style=hadamard edge] (77) to (89);
		\draw [style=hadamard edge] (73) to (85);
		\draw [style=hadamard edge] (73) to (89);
		\draw [style=hadamard edge] (81) to (85);
		\draw [style=hadamard edge] (81) to (89);
	\end{pgfonlayer}
\end{tikzpicture}
  \label{eqn:pivoting}
\end{equation}
These rewrites are useful as they reduce the number of vertices.
For a ZX graph with only Clifford vertices, successive application of these rules removes all spiders, reducing the diagram to a single scalar term.
This scalar could also be obtained by numerical tensor contraction of the underlying tensor network.
However, ZX contraction of a Clifford graph scales only cubically in the number of vertices, whereas tensor contraction can scale exponentially in the number of qubits.

\subsection{Doubled ZX notation}
\label{sec:doubling}
To represent measurements, we use the \emph{doubled} notation of the ZX calculus~\cite{van2020zx}. This notation introduces quantum vertices and wires, drawn in bold. A bold diagram denotes two independent copies with conjugated phases:
\begin{equation}
  \begin{tikzpicture}
	\begin{pgfonlayer}{nodelayer}
		\node [style=Z phase dot] (0) at (0, 0.5) {$\alpha$};
		\node [style=none] (1) at (1.75, 1.5) {};
		\node [style=none] (2) at (-1.75, 1.5) {};
		\node [style=none] (3) at (-1.75, -1) {};
		\node [style=none] (4) at (1.75, -1) {};
		\node [style=none] (5) at (1.75, 0.5) {};
		\node [style=none] (6) at (-1.75, 0.5) {};
		\node [style=none, rotate=90] (7) at (-1.5, -0.25) {...};
		\node [style=none, rotate=90] (8) at (1.5, -0.25) {...};
		\node [style=Z phase dot] (9) at (0, -0.5) {$\ -\alpha~$};
		\node [style=none] (10) at (1.75, 1.325) {};
		\node [style=none] (11) at (-1.75, 1.325) {};
		\node [style=none] (12) at (-1.75, -1.175) {};
		\node [style=none] (13) at (1.75, -1.175) {};
		\node [style=none] (14) at (1.75, 0.325) {};
		\node [style=none] (15) at (-1.75, 0.325) {};
		\node [style=Z phase dot bb] (16) at (-5, 0.25) {$\alpha$};
		\node [style=none] (17) at (-3.75, 1.5) {};
		\node [style=none] (18) at (-6.25, 1.5) {};
		\node [style=none] (19) at (-6.25, -1.25) {};
		\node [style=none] (20) at (-3.75, -1.25) {};
		\node [style=none] (21) at (-3.75, 0.5) {};
		\node [style=none] (22) at (-6.25, 0.5) {};
		\node [style=none, rotate=90] (23) at (-6, -0.25) {...};
		\node [style=none, rotate=90] (24) at (-4, -0.25) {...};
		\node [style=none] (181) at (-2.75, 0.25) {$=$};
	\end{pgfonlayer}
	\begin{pgfonlayer}{edgelayer}
		\draw [in=-141, out=0, looseness=0.75] (3.center) to (0);
		\draw [in=180, out=-39, looseness=0.75] (0) to (4.center);
		\draw [in=180, out=22, looseness=0.75] (0) to (5.center);
		\draw [in=180, out=39, looseness=0.75] (0) to (1.center);
		\draw [in=0, out=158, looseness=0.75] (0) to (6.center);
		\draw [in=141, out=0, looseness=0.75] (2.center) to (0);
		\draw [in=-141, out=0, looseness=0.75] (12.center) to (9);
		\draw [in=180, out=-39, looseness=0.75] (9) to (13.center);
		\draw [in=180, out=22, looseness=0.75] (9) to (14.center);
		\draw [in=180, out=39, looseness=0.75] (9) to (10.center);
		\draw [in=0, out=158, looseness=0.75] (9) to (15.center);
		\draw [in=141, out=0, looseness=0.75] (11.center) to (9);
		\draw [style=boldedge, in=-141, out=0, looseness=0.75] (19.center) to (16);
		\draw [style=boldedge, in=180, out=-39, looseness=0.75] (16) to (20.center);
		\draw [style=boldedge, in=180, out=22, looseness=0.75] (16) to (21.center);
		\draw [style=boldedge, in=180, out=39, looseness=0.75] (16) to (17.center);
		\draw [style=boldedge, in=0, out=158, looseness=0.75] (16) to (22.center);
		\draw [style=boldedge, in=141, out=0, looseness=0.75] (18.center) to (16);
	\end{pgfonlayer}
\end{tikzpicture}
  \label{eqn:doubling}
\end{equation}
To represent the classical flow of information, we make use of mixed diagrams:
\begin{equation}
  \begin{tikzpicture}
	\begin{pgfonlayer}{nodelayer}
		\node [style=Z dot] (16) at (-5, 0.25) {};
		\node [style=none] (17) at (-3.75, 1.5) {};
		\node [style=none] (18) at (-6.25, 1.5) {};
		\node [style=none] (19) at (-6.25, -1.25) {};
		\node [style=none] (20) at (-3.75, -1.25) {};
		\node [style=none] (21) at (-3.75, 0.5) {};
		\node [style=none] (22) at (-6.25, 0.5) {};
		\node [style=none, rotate=90] (23) at (-6, -0.25) {...};
		\node [style=none, rotate=90] (24) at (-4, -0.25) {...};
		\node [style=none] (181) at (-2.75, 0.25) {$=$};
		\node [style=Z dot] (182) at (-0.5, 0.25) {};
		\node [style=none] (183) at (0.75, 1.5) {};
		\node [style=none] (184) at (-1.75, 1.5) {};
		\node [style=none] (185) at (-1.75, -1) {};
		\node [style=none] (186) at (0.75, -1.25) {};
		\node [style=none] (187) at (0.75, 0.5) {};
		\node [style=none] (188) at (-1.75, 0.5) {};
		\node [style=none, rotate=90] (189) at (-1.5, -0.25) {...};
		\node [style=none, rotate=90] (190) at (0.5, -0.25) {...};
		\node [style=none] (191) at (-1.75, -1.25) {};
		\node [style=none] (192) at (-1.75, 1.25) {};
		\node [style=none] (193) at (-1.75, 0.25) {};
	\end{pgfonlayer}
	\begin{pgfonlayer}{edgelayer}
		\draw [style=boldedge, in=-141, out=0, looseness=0.75] (19.center) to (16);
		\draw [in=180, out=-39, looseness=0.75] (16) to (20.center);
		\draw [looseness=0.75, bend left=15] (16) to (21.center);
		\draw [in=180, out=39, looseness=0.75] (16) to (17.center);
		\draw [style=boldedge, in=0, out=158, looseness=0.75] (16) to (22.center);
		\draw [style=boldedge, in=141, out=0, looseness=0.75] (18.center) to (16);
		\draw [bend right=15] (185.center) to (182);
		\draw [in=180, out=-39, looseness=0.75] (182) to (186.center);
		\draw [looseness=0.75, bend left=15] (182) to (187.center);
		\draw [in=180, out=39, looseness=0.75] (182) to (183.center);
		\draw [looseness=0.75, bend left=15] (182) to (188.center);
		\draw [in=141, out=0, looseness=0.75] (184.center) to (182);
		\draw [out=-15, in=150] (192.center) to (182);
		\draw [bend right=15] (193.center) to (182);
		\draw [out=15, in=240, looseness=0.75] (191.center) to (182);
	\end{pgfonlayer}
\end{tikzpicture}
  \label{eqn:doubling2}
\end{equation}
With this, Z measurements are represented as
\begin{equation}
  \begin{tikzpicture}
	\begin{pgfonlayer}{nodelayer}
		\node [style=Z dot] (0) at (0, 0.25) {};
		\node [style=Z dot] (16) at (-5, 0.25) {};
		\node [style=none] (181) at (-2.5, 0.25) {$=$};
		\node [style=none] (183) at (-3.75, 0.25) {};
		\node [style=none] (184) at (-6.25, 0.25) {};
		\node [style=none] (185) at (1, 1) {};
		\node [style=none] (186) at (-1.25, 0.5) {};
		\node [style=none] (188) at (-1.25, 0) {};
		\node [style=none] (189) at (-4, 1) {};
		\node [style=none] (190) at (1.25, 0.5) {};
		\node [style=none] (191) at (1.25, 0) {};
	\end{pgfonlayer}
	\begin{pgfonlayer}{edgelayer}
		\draw [style=boldedge] (16) to (183.center);
		\draw [bend left=15] (186.center) to (0);
		\draw [bend right=15] (188.center) to (0);
		\draw [bend right=15] (185.center) to (0);
		\draw [style=boldedge] (184.center) to (16);
		\draw [bend left=15] (16) to (189.center);
		\draw [bend left=15] (0) to (190.center);
		\draw [bend right=15] (0) to (191.center);
	\end{pgfonlayer}
\end{tikzpicture}
  \label{eqn:measurement}
\end{equation}
where the classical wire carries the outcome of the measurement.
The measurement operation can be understood by decomposing the diagram as
\begin{equation}
  \begin{tikzpicture}
	\begin{pgfonlayer}{nodelayer}
		\node [style=Z dot] (0) at (0, 0.25) {};
		\node [style=none] (181) at (2.5, 0.25) {$=$};
		\node [style=none] (185) at (1, 2.25) {};
		\node [style=none] (186) at (-1.25, 0.75) {};
		\node [style=none] (188) at (-1.25, -0.25) {};
		\node [style=none] (190) at (1.25, 0.75) {};
		\node [style=none] (191) at (1.25, -0.25) {};
		\node [style=none] (194) at (9.5, 0.75) {};
		\node [style=X phase dot] (198) at (10.5, 0.75) {$\pi$};
		\node [style=none] (199) at (9.5, -0.25) {};
		\node [style=X phase dot] (200) at (10.5, -0.25) {$\pi$};
		\node [style=none] (201) at (13.5, 0.75) {};
		\node [style=X phase dot] (202) at (12.5, 0.75) {$\pi$};
		\node [style=none] (203) at (13.5, -0.25) {};
		\node [style=X phase dot] (204) at (12.5, -0.25) {$\pi$};
		\node [style=none] (205) at (12.5, 2.25) {};
		\node [style=X phase dot] (206) at (11.5, 1.5) {$\pi$};
		\node [style=none] (207) at (3.5, 0.75) {};
		\node [style=X phase dot] (208) at (4.5, 0.75) {$0$};
		\node [style=none] (209) at (3.5, -0.25) {};
		\node [style=X phase dot] (210) at (4.5, -0.25) {$0$};
		\node [style=none] (211) at (7.5, 0.75) {};
		\node [style=X phase dot] (212) at (6.5, 0.75) {$0$};
		\node [style=none] (213) at (7.5, -0.25) {};
		\node [style=X phase dot] (214) at (6.5, -0.25) {$0$};
		\node [style=none] (215) at (6.5, 2.25) {};
		\node [style=X phase dot] (216) at (5.5, 1.5) {$0$};
		\node [style=none] (217) at (8.5, 0.25) {$+$};
		\node [style=none] (218) at (-2, 0.25) {$\sqrt{2}$};
	\end{pgfonlayer}
	\begin{pgfonlayer}{edgelayer}
		\draw [bend left=15] (186.center) to (0);
		\draw [bend right=15] (188.center) to (0);
		\draw [bend right=15] (185.center) to (0);
		\draw [bend left=15] (0) to (190.center);
		\draw [bend right=15] (0) to (191.center);
		\draw (194.center) to (198);
		\draw (199.center) to (200);
		\draw (201.center) to (202);
		\draw (203.center) to (204);
		\draw [out=-165, in=60] (205.center) to (206);
		\draw (207.center) to (208);
		\draw (209.center) to (210);
		\draw (211.center) to (212);
		\draw (213.center) to (214);
		\draw [out=-165, in=60] (215.center) to (216);
	\end{pgfonlayer}
\end{tikzpicture}
  \label{eqn:measurement_cutting}
\end{equation}
making it clear that the channel projects onto and re-initializes in either the $|0\rangle$ or $|1\rangle$ state, with the measurement outcome carried by a disconnected X spider.

When a measurement result is not recorded, we obtain the trace. The ZX diagrams for discarding a Z or X measurement are identical, as a direct consequence of causality:
\begin{equation}
  \begin{tikzpicture}
	\begin{pgfonlayer}{nodelayer}
		\node [style=Z dot] (16) at (-5, 0.25) {};
		\node [style=none] (181) at (-4, 0.25) {$=$};
		\node [style=none] (184) at (-6.25, 0.25) {};
		\node [style=none] (186) at (1.25, 0.5) {};
		\node [style=none] (188) at (1.25, 0) {};
		\node [style=X dot] (189) at (-1.75, 0.25) {};
		\node [style=none] (190) at (-3, 0.25) {};
		\node [style=none] (191) at (-0.75, 0.25) {$=$};
		\node [style=none] (192) at (0.25, 0.5) {};
		\node [style=none] (193) at (0.25, 0) {};
	\end{pgfonlayer}
	\begin{pgfonlayer}{edgelayer}
		\draw [style=boldedge] (184.center) to (16);
		\draw [style=boldedge] (190.center) to (189);
		\draw [looseness=1.75, bend left=90] (186.center) to (188.center);
		\draw (192.center) to (186.center);
		\draw (193.center) to (188.center);
	\end{pgfonlayer}
\end{tikzpicture}
  \label{eqn:trace}
\end{equation}

\subsection{Pauli noise channels}
\label{sec:pauli_channels}
Pauli noise channels are represented as parameterized X and Z vertices with phase in $[0,\pi]$.
X, Y, or Z bit flip channels are represented by the following ZX diagrams:
\begin{equation}
  \begin{tikzpicture}
	\begin{pgfonlayer}{nodelayer}
		\node [style=X phase dot bb] (1) at (-3.75, -0.25) {$\;e_i\pi\;$};
		\node [style=none] (3) at (-5, -0.25) {};
		\node [style=none] (4) at (-2.5, -0.25) {};
		\node [style=X phase dot bb] (7) at (0, -0.25) {$\;e_i\pi\;$};
		\node [style=Z phase dot bb] (8) at (1.5, -0.25) {$\;e_i\pi\;$};
		\node [style=none] (9) at (-1.25, -0.25) {};
		\node [style=none] (10) at (2.75, -0.25) {};
		\node [style=Z phase dot bb] (12) at (5.25, -0.25) {$\;e_i\pi\;$};
		\node [style=none] (13) at (4, -0.25) {};
		\node [style=none] (14) at (6.5, -0.25) {};
	\end{pgfonlayer}
	\begin{pgfonlayer}{edgelayer}
		\draw [style=boldedge] (3.center) to (1);
		\draw [style=boldedge] (9.center) to (7);
		\draw [style=boldedge] (7) to (8);
		\draw [style=boldedge] (8) to (10.center);
		\draw [style=boldedge] (12) to (14.center);
		\draw [style=boldedge] (1) to (4.center);
		\draw [style=boldedge] (13.center) to (12);
	\end{pgfonlayer}
\end{tikzpicture}
  \label{eqn:channel_flip}
\end{equation}
Here, the binary variable $e_i$ is sampled from the channel's probability distribution.
A general single-qubit Pauli channel
\begin{equation}
  \begin{tikzpicture}
	\begin{pgfonlayer}{nodelayer}
		\node [style=X phase dot bb] (7) at (0, -0.25) {$\;e_i\pi\;$};
		\node [style=Z phase dot bb] (8) at (1.5, -0.25) {$\;e_j\pi\;$};
		\node [style=none] (9) at (-1.25, -0.25) {};
		\node [style=none] (10) at (2.75, -0.25) {};
	\end{pgfonlayer}
	\begin{pgfonlayer}{edgelayer}
		\draw [style=boldedge] (9.center) to (7);
		\draw [style=boldedge] (7) to (8);
		\draw [style=boldedge] (8) to (10.center);
	\end{pgfonlayer}
\end{tikzpicture}
  \label{eqn:channel_pauli1}
\end{equation}
requires two random bits, where configuration $(e_i=1,e_j=0)$ is sampled with probability $p_x$, $(1,1)$ with probability $p_y$, $(0,1)$ with $p_z$, and $(0,0)$ with probability $1-p_x-p_y-p_z$.
Multi-qubit Pauli channels are constructed analogously, e.g., with the diagram
\begin{equation}
  \begin{tikzpicture}
	\begin{pgfonlayer}{nodelayer}
		\node [style=X phase dot bb] (7) at (0, -0.25) {$\;e_i\pi\;$};
		\node [style=Z phase dot bb] (8) at (1.5, -0.25) {$\;e_j\pi\;$};
		\node [style=none] (9) at (-1.25, -0.25) {};
		\node [style=none] (10) at (2.75, -0.25) {};
		\node [style=X phase dot bb] (11) at (0, -1.25) {$\;e_k\pi\;$};
		\node [style=Z phase dot bb] (12) at (1.5, -1.25) {$\;e_l\pi\;$};
		\node [style=none] (13) at (-1.25, -1.25) {};
		\node [style=none] (14) at (2.75, -1.25) {};
	\end{pgfonlayer}
	\begin{pgfonlayer}{edgelayer}
		\draw [style=boldedge] (9.center) to (7);
		\draw [style=boldedge] (7) to (8);
		\draw [style=boldedge] (8) to (10.center);
		\draw [style=boldedge] (13.center) to (11);
		\draw [style=boldedge] (11) to (12);
		\draw [style=boldedge] (12) to (14.center);
	\end{pgfonlayer}
\end{tikzpicture}
  \label{eqn:channel_pauli2}
\end{equation}
which requires classical sampling of four bits $e_i,e_j,e_k,e_l$.

\section{Channel reduction}
\label{sec:channel_reduction}

Each noise channel in the circuit defines a probability distribution $p(e_{i_1},\dots,e_{i_n})$ over a group of error bits that share a joint distribution. These bits correspond to columns in the binary matrix $T_{ij}$. The goal of channel reduction is to minimize the number of error bits that must be sampled per shot, without altering the induced distribution over detectors and observables.

Tsim applies four successive transformations. First, \emph{null-column elimination} identifies error bits whose corresponding columns in $T_{ij}$ are identically zero, meaning they do not flip any detector or observable regardless of their value. These bits are marginalized out of their respective channels by summing over them, and channels that consist entirely of null bits are removed altogether.

Second, \emph{channel normalization} sorts the column indices within each channel into a canonical order and permutes the probability array accordingly. This ensures that two channels affecting the same set of columns are represented identically, which is a prerequisite for the subsequent merging step.

Third, \emph{identical-channel merging} groups all channels that share the same column signature and combines each group into a single channel by XOR convolution. Given two independent channels $A$ and $B$ over the same set of error bits, the probability of the combined outcome $o$ is $P(o) = \sum_{a \oplus b = o} P(A{=}a)\,P(B{=}b)$. This reduces multiple independent noise sources with identical detector--observable flip patterns to a single effective channel.

Finally, \emph{subset absorption} identifies channels whose column signatures are strict subsets of another channel's signature. When channel $A$ acts on columns $\{c_1, c_2\}$ and channel $B$ acts on $\{c_1, c_2, c_3\}$, channel $A$ is expanded to the larger signature by embedding its distribution into the higher-dimensional space (assigning zero probability to outcomes involving the new bits) and then convolved with $B$. Channels are processed in decreasing order of signature size so that larger channels absorb smaller ones.

Note that if all joint probability distributions are first factorized into approximate, independent single-bit channels, the above reduction pipeline yields a set of independent error mechanisms, each characterized by a single probability and a column signature in $T_{ij}$. This is precisely the detector error model~\cite{gidney2021stim}, which serves as the standard input for decoders.

\section{Stabilizer decompositions}
\label{sec:decomp}

So far, we have described the evaluation of parameterized Clifford ZX diagrams.
The ZX reduction rules (\ref{eqn:local_comp}, \ref{eqn:pivoting}) terminate when there are no internal $\pi/2$-vertices and no pairs
of $\pi$-vertices. For the Clifford segment, this is sufficient to remove all vertices
and obtain a single scalar term. However, when non-Clifford vertices are present, this
routine generally does not produce a scalar diagram. Instead, one obtains diagrams
with non-Clifford vertices and unpaired $\pi$-vertices.

To proceed, we decompose non-Clifford vertices into a sum of Clifford
vertices. For example, a T state can be decomposed into a sum of two Clifford terms:
\begin{equation}
  \begin{tikzpicture}
	\begin{pgfonlayer}{nodelayer}
		\node [style=Z phase dot] (289) at (4.75, 0) {$\frac{\pi}{4}$};
		\node [style=none] (291) at (5.75, 0) {};
		\node [style=none] (292) at (6.75, 0) {$=$};
		\node [style=none] (285) at (11.25, 0) {$+$};
		\node [style=X dot] (293) at (9.25, 0) {};
		\node [style=none] (294) at (10.25, 0) {};
		\node [style=X phase dot] (295) at (15, 0) {$\pi$};
		\node [style=none] (296) at (16.25, 0) {};
		\node [style=none] (297) at (8, 0) {$\frac{1}{\sqrt{2}}$};
		\node [style=none] (298) at (13, 0) {$\frac{1}{\sqrt{2}}e^{i\pi /4}$};
	\end{pgfonlayer}
	\begin{pgfonlayer}{edgelayer}
		\draw (289) to (291.center);
		\draw (293) to (294.center);
		\draw (295) to (296.center);
	\end{pgfonlayer}
\end{tikzpicture}
\end{equation}
Since one can always unfuse non-Clifford vertices, each $\pi/4$-vertex can be removed at the cost of doubling the number of diagrams. Naively, decomposing $n$ non-Clifford vertices generates $2^n$ Clifford diagrams, but more favorable identities exist when multiple vertices are decomposed simultaneously. For example, two T-gates can be decomposed as:
\begin{equation}
  \begin{tikzpicture}
	\begin{pgfonlayer}{nodelayer}
		\node [style=Z phase dot] (0) at (-1.5, 0.488) {$\frac\pi4$};
		\node [style=none] (1) at (-0.475, 0.487) {};
		\node [style=Z phase dot] (2) at (-1.475, -0.512) {$\frac\pi4$};
		\node [style=none] (3) at (-0.45, -0.513) {};
		\node [style=none] (4) at (0.5, 0) {$=$};
		\node [style=Z phase dot] (5) at (1.75, -0.012) {$\frac\pi2$};
		\node [style=none] (6) at (2.775, 0.487) {};
		\node [style=none] (8) at (2.8, -0.513) {};
		\node [style=none] (9) at (4, 0) {$+\ e^{i\frac\pi4}$};
		\node [style=X phase dot] (15) at (5.75, -0.01200000000000001) {$\pi$};
		\node [style=none] (16) at (6.775, 0.487) {};
		\node [style=none] (17) at (6.8, -0.5129999999999999) {};
	\end{pgfonlayer}
	\begin{pgfonlayer}{edgelayer}
		\draw (1.center) to (0);
		\draw (3.center) to (2);
		\draw [bend right] (6.center) to (5);
		\draw [bend right] (5) to (8.center);
		\draw [bend right] (16.center) to (15);
		\draw [bend right] (15) to (17.center);
	\end{pgfonlayer}
\end{tikzpicture}
\end{equation}
This leads to exponential growth with $2^{\alpha n}$, where $\alpha=0.5$. By default, Tsim uses the partial \textit{Cat5} decomposition first proposed in~\cite{qassim2021improved}:
\begin{equation}
\label{eqn:cat5}
  \begin{tikzpicture}
	\begin{pgfonlayer}{nodelayer}
		\node [style=none] (98) at (-8.45, 2.25) {$=$};
		\node [style=none] (120) at (-6.4, 2.25) {$\left(1 - e^{4i \theta}\right)$};
		\node [style=Z phase dot] (199) at (-10.25, 4.25) {$\theta$};
		\node [style=none] (204) at (-9.25, 4.25) {};
		\node [style=Z phase dot] (205) at (-10.25, 3.25) {$\theta$};
		\node [style=none] (206) at (-9.25, 3.25) {};
		\node [style=Z phase dot] (207) at (-10.25, 2.25) {$\theta$};
		\node [style=none] (208) at (-9.25, 2.25) {};
		\node [style=Z phase dot] (209) at (-10.25, 1.25) {$\theta$};
		\node [style=none] (210) at (-9.25, 1.25) {};
		\node [style=Z phase dot] (211) at (-10.25, 0.25) {$\theta$};
		\node [style=none] (212) at (-9.25, 0.25) {};
		\node [style=none] (223) at (-1.2000000000000002, 2.25) {$+$};
		\node [style=Z dot] (234) at (7, 4.25) {};
		\node [style=none] (235) at (8, 4.25) {};
		\node [style=Z dot] (236) at (7, 3.25) {};
		\node [style=none] (237) at (8, 3.25) {};
		\node [style=Z dot] (238) at (7, 2.25) {};
		\node [style=none] (239) at (8, 2.25) {};
		\node [style=Z dot] (240) at (7, 1.25) {};
		\node [style=none] (241) at (8, 1.25) {};
		\node [style=Z dot] (242) at (7, 0.25) {};
		\node [style=none] (243) at (8, 0.25) {};
		\node [style=Z dot] (244) at (5.5, 2.25) {};
		\node [style=Z phase dot] (245) at (3.75, 2.25) {$\;-\theta\;$};
		\node [style=Z phase dot] (246) at (17.5, 4.25) {$\frac{\pi}{2}$};
		\node [style=none] (247) at (18.5, 4.25) {};
		\node [style=Z phase dot] (248) at (17.5, 3.25) {$\frac{\pi}{2}$};
		\node [style=none] (249) at (18.5, 3.25) {};
		\node [style=Z phase dot] (250) at (17.5, 2.25) {$\frac{\pi}{2}$};
		\node [style=none] (251) at (18.5, 2.25) {};
		\node [style=Z phase dot] (252) at (17.5, 1.25) {$\frac{\pi}{2}$};
		\node [style=none] (253) at (18.5, 1.25) {};
		\node [style=Z phase dot] (254) at (17.5, 0.25) {$\frac{\pi}{2}$};
		\node [style=none] (255) at (18.5, 0.25) {};
		\node [style=Z dot] (256) at (16, 2.25) {};
		\node [style=Z phase dot] (257) at (14.25, 2.25) {$\;\frac{\pi}{2}-\theta\;$};
		\node [style=none] (260) at (1.0999999999999996, 2.25) {$2\,e^{3i\theta}\cos \theta$};
		\node [style=none] (261) at (8.8, 2.25) {$+$};
		\node [style=none] (262) at (11.100000000000001, 2.25) {$2\,e^{3i\theta}i \sin \theta$};
		\node [style=none] (263) at (-2, 4.25) {};
		\node [style=none] (265) at (-2, 3.25) {};
		\node [style=none] (267) at (-2, 2.25) {};
		\node [style=none] (269) at (-2, 1.25) {};
		\node [style=none] (271) at (-2, 0.25) {};
		\node [style=Z phase dot] (273) at (-3.75, 2.25) {$\;\theta+\pi\;$};
	\end{pgfonlayer}
	\begin{pgfonlayer}{edgelayer}
		\draw (199) to (204.center);
		\draw (205) to (206.center);
		\draw (207) to (208.center);
		\draw (209) to (210.center);
		\draw (211) to (212.center);
		\draw (234) to (235.center);
		\draw (236) to (237.center);
		\draw (238) to (239.center);
		\draw (240) to (241.center);
		\draw (242) to (243.center);
		\draw [style=hadamard edge] (244) to (238);
		\draw [style=hadamard edge, bend left=15] (244) to (236);
		\draw [style=hadamard edge, bend left=15] (244) to (234);
		\draw [style=hadamard edge, bend right=15] (244) to (240);
		\draw [style=hadamard edge, bend right=15] (244) to (242);
		\draw [style=hadamard edge] (244) to (245);
		\draw (246) to (247.center);
		\draw (248) to (249.center);
		\draw (250) to (251.center);
		\draw (252) to (253.center);
		\draw (254) to (255.center);
		\draw [style=hadamard edge] (256) to (250);
		\draw [style=hadamard edge, bend left=15] (256) to (248);
		\draw [style=hadamard edge, bend left=15] (256) to (246);
		\draw [style=hadamard edge, bend right=15] (256) to (252);
		\draw [style=hadamard edge, bend right=15] (256) to (254);
		\draw [style=hadamard edge] (256) to (257);
		\draw (273) to (267.center);
		\draw [looseness=1.25, bend left=15] (273) to (265.center);
		\draw [bend left] (273) to (263.center);
		\draw [bend right=15] (273) to (269.center);
		\draw [bend right] (273) to (271.center);
	\end{pgfonlayer}
\end{tikzpicture}
\end{equation}
Importantly, this decomposition applies to arbitrary rotation angles $\theta$. For $\theta=\frac{\pi}{4}$, it reduces to the ZX diagram previously reported
in~\cite{kissinger2022classical}.
In the asymptotic limit, the exponential growth rate is $\alpha\approx 0.396$.

\section{Sampling from marginals}

Marginal probability distributions are represented by diagrams where measurements are traced out. For example, $P(m_0,\dots,m_{k-1})$ is given by:
\begin{equation}
  \begin{tikzpicture}
	\begin{pgfonlayer}{nodelayer}
		\node [style=X dot bb] (32) at (-9.5, 2.25) {};
		\node [style=Z dot] (33) at (-2.25, 2.25) {};
		\node [style=X dot bb] (34) at (-9.5, 0.75) {};
		\node [style=Z dot] (35) at (-2.25, 0.75) {};
		\node [style=X dot bb] (36) at (-9.5, -0.75) {};
		\node [style=Z dot] (37) at (-2.25, -0.75) {};
		\node [style=X dot bb] (53) at (-9.5, -2.75) {};
		\node [style=Z dot] (54) at (-2.25, -2.75) {};
		\node [style=none] (55) at (-7.5, 2.75) {};
		\node [style=none] (56) at (-7.5, -3.25) {};
		\node [style=none] (57) at (-3.5, -3.25) {};
		\node [style=none] (58) at (-3.5, 2.75) {};
		\node [style=none] (59) at (-7.5, 2.25) {};
		\node [style=none] (60) at (-7.5, 0.75) {};
		\node [style=none] (61) at (-7.5, -0.75) {};
		\node [style=none] (63) at (-7.5, -2.75) {};
		\node [style=none] (64) at (-3.5, -2.75) {};
		\node [style=none] (66) at (-3.5, -0.75) {};
		\node [style=none] (67) at (-3.5, 0.75) {};
		\node [style=none] (68) at (-3.5, 2.25) {};
		\node [style=none] (69) at (-5.5, -0.125) {$U$};
		\node [style=none, rotate=90] (168) at (-8.25, -1.75) {$\cdots$};
		\node [style=none, rotate=90] (169) at (-2.75, -1.75) {$\cdots$};
		\node [style=none, rotate=90] (170) at (-2.75, 1.5) {$\cdots$};
		\node [style=none, rotate=90] (195) at (-8.25, 1.5) {$\cdots$};
		\node [style=X phase dot] (196) at (-1, 2.25) {$\;m_0\pi\;$};
		\node [style=X phase dot] (197) at (-1, 0.75) {$\;m_{k}\pi\;$};
		\node [style=none] (215) at (-12.5, 0) {$P(m_0 \ldots m_{k})\;=$};
	\end{pgfonlayer}
	\begin{pgfonlayer}{edgelayer}
		\draw [style=boldedge] (55.center) to (58.center);
		\draw [style=boldedge] (58.center) to (57.center);
		\draw [style=boldedge] (57.center) to (56.center);
		\draw [style=boldedge] (56.center) to (55.center);
		\draw [style=boldedge] (68.center) to (33);
		\draw [style=boldedge] (53) to (63.center);
		\draw [style=boldedge] (36) to (61.center);
		\draw [style=boldedge] (60.center) to (34);
		\draw [style=boldedge] (32) to (59.center);
		\draw [style=boldedge] (64.center) to (54);
		\draw [style=boldedge] (37) to (66.center);
		\draw [style=boldedge] (67.center) to (35);
		\draw (33) to (196);
		\draw (35) to (197);
		\draw (215.center) to (215.center);
	\end{pgfonlayer}
\end{tikzpicture}
  \label{eqn:zx_marginal_sum}
\end{equation}
We sample bit strings from the circuit's output distribution via autoregressive conditioning on marginal probabilities~\cite{kissinger2022classical,kissinger2022simulating}. The procedure is summarized in Algorithm~\ref{alg:autoregressive}.

\begin{algorithm}[H]
  \caption{Autoregressive sampling from marginal probabilities}\label{alg:autoregressive}
  \begin{algorithmic}[1]
    \Require Marginal probability functions $P$, $P(m_0)$, $P(m_0, m_1)$, ..., $P(m_0, \dots, m_{n-1})$ obtained from ZX diagrams
    \Ensure Sample $(\tilde m_0, \dots, \tilde m_{n-1})$ drawn from the circuit output distribution
    \State $P_{\mathrm{prev}} \gets P$ \Comment{Normalization}
    \For{$i = 0, \dots, n-1$}
    \State $P_1 \gets P(m_0{=}\tilde m_0,\;\dots,\;m_{i-1}{=}\tilde m_{i-1},\;m_i{=}1)$
    \State Draw $\tilde m_i \sim \mathrm{Bernoulli}(P_1 / P_{\mathrm{prev}})$
    \If{$\tilde m_i = 1$}
    \State $P_{\mathrm{prev}} \gets P_1$
    \Else
    \State $P_{\mathrm{prev}} \gets P_{\mathrm{prev}} - P_1$
    \EndIf
    \EndFor
    \State \Return $(\tilde m_0, \dots, \tilde m_{n-1})$
  \end{algorithmic}
\end{algorithm}

\subsection{Sampling detectors and observables}
\label{sec:detectors_observables}

In QEC experiments, one is usually not directly interested in measurements but rather in a smaller set of detectors and observables. Detectors and observables correspond to linear XOR transformations of measurements. For ZX graphs, XOR transformations have a simple representation:
\begin{equation}
  \begin{tikzpicture}
	\begin{pgfonlayer}{nodelayer}
		\node [style=X dot] (0) at (0, 0.25) {};
		\node [style=none] (185) at (1.25, 0.25) {};
		\node [style=none] (186) at (-1.25, 1) {};
		\node [style=none] (188) at (-1.25, -0.5) {};
		\node [style=none] (189) at (-1.25, 0.25) {};
		\node [style=none] (190) at (-2, 1) {$m_i$};
		\node [style=none] (191) at (-2, -0.5) {$m_k$};
		\node [style=none] (192) at (-2, 0.25) {$m_j$};
		\node [style=none] (193) at (3.75, 0.25) {$m_i \oplus m_j \oplus m_k$};
	\end{pgfonlayer}
	\begin{pgfonlayer}{edgelayer}
		\draw [bend left=15] (186.center) to (0);
		\draw [bend right=15] (188.center) to (0);
		\draw (185.center) to (0);
		\draw (189.center) to (0);
		\draw (186.center) to (186.center);
		\draw (190.center) to (190.center);
		\draw (193.center) to (193.center);
	\end{pgfonlayer}
\end{tikzpicture}
  \label{eqn:xor}
\end{equation}
Detector outcomes are then sampled by computing their marginal probabilities, e.g., by evaluating diagrams of the form:
\begin{equation}
  \begin{tikzpicture}
	\begin{pgfonlayer}{nodelayer}
		\node [style=X dot bb] (32) at (-9.5, 2.25) {};
		\node [style=Z dot] (33) at (-2.25, 2.25) {};
		\node [style=X dot bb] (34) at (-9.5, 0.75) {};
		\node [style=Z dot] (35) at (-2.25, 0.75) {};
		\node [style=X dot bb] (36) at (-9.5, -0.75) {};
		\node [style=Z dot] (37) at (-2.25, -0.75) {};
		\node [style=X dot bb] (53) at (-9.5, -2.75) {};
		\node [style=Z dot] (54) at (-2.25, -2.75) {};
		\node [style=none] (55) at (-7.5, 2.75) {};
		\node [style=none] (56) at (-7.5, -3.25) {};
		\node [style=none] (57) at (-3.5, -3.25) {};
		\node [style=none] (58) at (-3.5, 2.75) {};
		\node [style=none] (59) at (-7.5, 2.25) {};
		\node [style=none] (60) at (-7.5, 0.75) {};
		\node [style=none] (61) at (-7.5, -0.75) {};
		\node [style=none] (63) at (-7.5, -2.75) {};
		\node [style=none] (64) at (-3.5, -2.75) {};
		\node [style=none] (66) at (-3.5, -0.75) {};
		\node [style=none] (67) at (-3.5, 0.75) {};
		\node [style=none] (68) at (-3.5, 2.25) {};
		\node [style=none] (69) at (-5.5, -0.125) {$U$};
		\node [style=none, rotate=90] (168) at (-8.25, -1.75) {$\cdots$};
		\node [style=none, rotate=90] (169) at (-2.75, -1.75) {$\cdots$};
		\node [style=none, rotate=90] (170) at (-2.75, 1.5) {$\cdots$};
		\node [style=none, rotate=90] (195) at (-8.25, 1.5) {$\cdots$};
		\node [style=X dot] (196) at (-1, 4.5) {};
		\node [style=X dot] (197) at (0.5, 4.5) {};
		\node [style=X dot] (198) at (2, 4.5) {};
		\node [style=X dot] (199) at (3.5, 4.5) {};
		\node [style=X phase dot] (200) at (-1, 5.75) {$\;d_0\pi\;$};
		\node [style=X phase dot] (201) at (0.5, 5.75) {$\;d_k\pi\;$};
		\node [style=Z dot] (202) at (2, 5.75) {};
		\node [style=Z dot] (203) at (3.5, 5.75) {};
		\node [style=none, rotate=90] (204) at (-0.5, 5) {$\vdots$};
		\node [style=none, rotate=90] (205) at (2.5, 5) {$\vdots$};
		\node [style=none] (210) at (-2.25, 2.25) {};
		\node [style=none] (211) at (-2.25, 0.75) {};
		\node [style=none] (212) at (-2.25, -0.75) {};
		\node [style=none] (213) at (-2.25, -2.75) {};
		\node [style=none] (215) at (-12.5, 0) {$P(d_0\dots d_k)\;=$};
	\end{pgfonlayer}
	\begin{pgfonlayer}{edgelayer}
		\draw [style=boldedge] (55.center) to (58.center);
		\draw [style=boldedge] (58.center) to (57.center);
		\draw [style=boldedge] (57.center) to (56.center);
		\draw [style=boldedge] (56.center) to (55.center);
		\draw [style=boldedge] (68.center) to (33);
		\draw [style=boldedge] (53) to (63.center);
		\draw [style=boldedge] (36) to (61.center);
		\draw [style=boldedge] (60.center) to (34);
		\draw [style=boldedge] (32) to (59.center);
		\draw [style=boldedge] (64.center) to (54);
		\draw [style=boldedge] (37) to (66.center);
		\draw [style=boldedge] (67.center) to (35);
		\draw (196) to (200);
		\draw (197) to (201);
		\draw (198) to (202);
		\draw (199) to (203);
		\draw (170.center) to (170.center);
		\draw (204.center) to (204.center);
		\draw (205.center) to (205.center);
		\draw [looseness=0.75, bend left] (196) to (210.center);
		\draw [looseness=0.5, bend left=15] (196) to (211.center);
		\draw [looseness=0.75, bend left=15] (197) to (211.center);
		\draw [looseness=0.75, bend left=15] (197) to (213.center);
		\draw [bend left=15] (198) to (212.center);
		\draw [looseness=0.5, bend left] (199) to (212.center);
		\draw [looseness=0.5, bend left=15] (199) to (211.center);
		\draw [looseness=0.5, bend left] (199) to (213.center);
	\end{pgfonlayer}
\end{tikzpicture}
  \label{eqn:zx_marginal_detectors}
\end{equation}

\section{Parameterized evaluation of Clifford ZX diagrams}
\label{sec:paramzx}

In the presence of parameterized Pauli vertices, simplification of a Clifford ZX diagram results in a product of phases, each depending on parities of a subset of binary parameters $f_i$.
Let us define parities of two specific subsets of the $f_i$ by
\begin{align}
  a &= f_{i_1} \oplus f_{i_2} \oplus \ldots \oplus f_{i_n} \\
  b &= f_{j_1} \oplus \ldots \oplus f_{j_m}\,.
\end{align}
In terms of these parities, diagrammatic reduction rules give rise to phase terms with the following structure:
\begin{table}[H]
  \centering
  {
    \renewcommand{\arraystretch}{2}
    \begin{tabular}{>{\centering\arraybackslash}p{3cm} >{\centering\arraybackslash}p{5cm} >{\centering\arraybackslash}p{4cm}}
      \textbf{Name} &
      \textbf{Form} &
      \textbf{Origin} \\ \hline\hline

      Node &
      $1+e^{i\left(\alpha +a\pi\right)}$ &
      (h) \\ \hline

      Half-$\pi$ &
      $e^{\pm i a \frac{\pi}{2}}$&
      (d.2), \eqref{eqn:local_comp} \\ \hline


      $\pi$-pair &
      $e^{i\pi a b} = (-1)^{ab}$ &
      \eqref{eqn:pivoting}, (e) \\ \hline

      Phase-pair &
      $1+e^{i (\alpha + a\pi) } + e^{i(\beta + b\pi)} - e^{i (\alpha + \beta +a \pi+b\pi)}$ &
      (b)  \\ \hline
    \end{tabular}
  }
  \caption{Types of parameterized phase terms arising from ZX reduction, where $a, b \in \{0, 1\}$ are parities of the form $a = f_{i_1} \oplus f_{i_2} \oplus \ldots \oplus f_{i_n}$ and $b = f_{j_1} \oplus \ldots \oplus f_{j_m}$.}
  \label{tab:paramzx:subtermtypes}
\end{table}

Formally, ZX reduction of a parameterized Clifford diagram yields a single scalar amplitude that is given, up to a global constant, by a product of phase factors from Table~\ref{tab:paramzx:subtermtypes}. Each phase factor depends on a subset of the binary parameters $f_i$.

Without loss of generality, we restrict to phase-pair terms, since all other term types can be represented as special cases of phase-pair terms~\cite{Sutcliffe_2025}. Tsim tracks individual phase terms for efficient evaluation.

Let $f_1, \ldots, f_n \in \{0,1\}$ denote the binary parameters. Each phase-pair term $k$ is associated with two selector vectors $u_{ki}, v_{ki} \in \{0,1\}$ that determine which parameters participate in its two parities:
\begin{equation}
  a_k = \bigoplus_{i} u_{ki} \, f_i \,,\qquad
  b_k = \bigoplus_{i} v_{ki} \, f_i \,.
  \label{eqn:parities}
\end{equation}
A single phase-pair term then reads
\begin{equation}
  h(\alpha_k, \beta_k; a_k, b_k)
  = 1 + e^{i(\alpha_k + a_k\pi)}
  + e^{i(\beta_k + b_k\pi)}
  - e^{i(\alpha_k + \beta_k + a_k\pi + b_k\pi)}\,,
  \label{eqn:phase_pair}
\end{equation}
where $\alpha_k, \beta_k$ are continuous phases fixed during diagram reduction.

The complete scalar resulting from ZX reduction is
\begin{equation}
  S(f_1,\ldots,f_n) = c \prod_{k=1}^{m} h(\alpha_k, \beta_k; a_k, b_k)\,,
  \label{eqn:scalar_product}
\end{equation}
where $c$ is a constant prefactor.

When non-Clifford vertices are present, the diagram must first be decomposed into a sum of Clifford diagrams, each of which reduces to a product of phase terms. The total amplitude is then

\begin{align}
  P = \sum_t S^{t}(\{f_i\}) =\sum_t c^{t} \prod_{k=1}^{m} h(\alpha_k^{t}, \beta_k^{t}; \bigoplus_{i} u_{ki}^t f_i, \bigoplus_{i} v_{ki}^t f_i)\,,
\end{align}

A parameterized ZX diagram is thus completely specified by the tensors
$c^t$, $\alpha_k^t$, $\beta_k^t$, $u_{ki}^t$, and $v_{ki}^t$,
where $i$ indexes $f$-variables, $k$ indexes phase terms, and $t$ indexes stabilizer rank decomposition terms.

For more efficient evaluation, especially on GPU, we introduce an additional batch dimension $l$ to compute multiple amplitudes with different parameter configurations $f_{il}$ in parallel:
\begin{align}
  P_l = \sum_t S^{t}(\{f_{il}\}) =\sum_t c^{t} \prod_{k=1}^{m} h(\alpha_k^{t}, \beta_k^{t}; \bigoplus_{i} u_{ki}^t f_{il}, \bigoplus_{i} v_{ki}^t f_{il})\,,
\end{align}
The computational bottleneck in evaluating $P_l$ is the tensor contraction $\bigoplus_{i} u_{ki}^t f_{il}$. Although $u_{ki}^t$
and $f_{il}$ are binary, Tsim converts them to floating-point format before contraction to utilize fast BLAS kernels.

\section{Validation}

\begin{figure}[t]
  \centering
  \includegraphics[width=\linewidth]{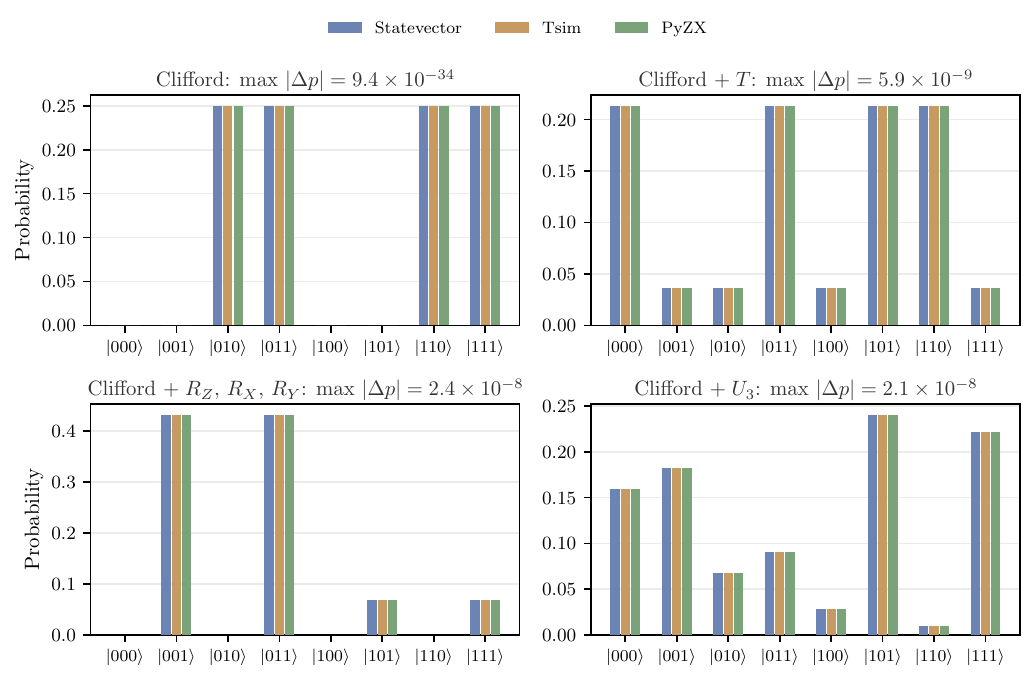}
  \caption{Validation of Tsim against a statevector simulator \cite{gidney2024magic} and PyZX tensor contraction for random circuits with terminal measurements.}
  \label{fig:validation}
\end{figure}

The Tsim repository includes a CI-backed test suite with linting, type checking (pyright), formatting (black), docstring checks, unit tests, and integration tests, with more than 95\% test coverage. Because Tsim is a sampling simulator and output distributions can be difficult to compare directly based on samples alone, Tsim provides functionality to compute probabilities of bit-strings. Fig.~\ref{fig:validation} validates these against a state-vector simulator~\cite{gidney2024magic} and PyZX tensor contraction on random circuits with terminal measurements, with agreement up to fp32 precision across randomized circuits of varying depth and gate type.

\section{Additional benchmarks}

CPU benchmarks were performed on an Apple M4 Pro. GPU benchmarks were performed on an NVIDIA Grace Hopper GH200 system with a 72-core ARM Neoverse-V2 host CPU and on an NVIDIA RTX 5090 GPU with AMD Ryzen 9 9900X.

For Tsim, we autotune the shot batch size separately for each benchmark instance: Starting from an initial batch size, we repeatedly double the batch size until the achieved throughput no longer improves or the requested batch no longer fits in device memory. On the GH200 system, the best throughput is typically obtained only at large batch sizes that make aggressive use of most of the available 96\,GB of VRAM on the GH200 or 32\,GB of VRAM on the RTX 5090.

\begin{figure}[t]
  \centering
  \includegraphics[width=0.85\linewidth]{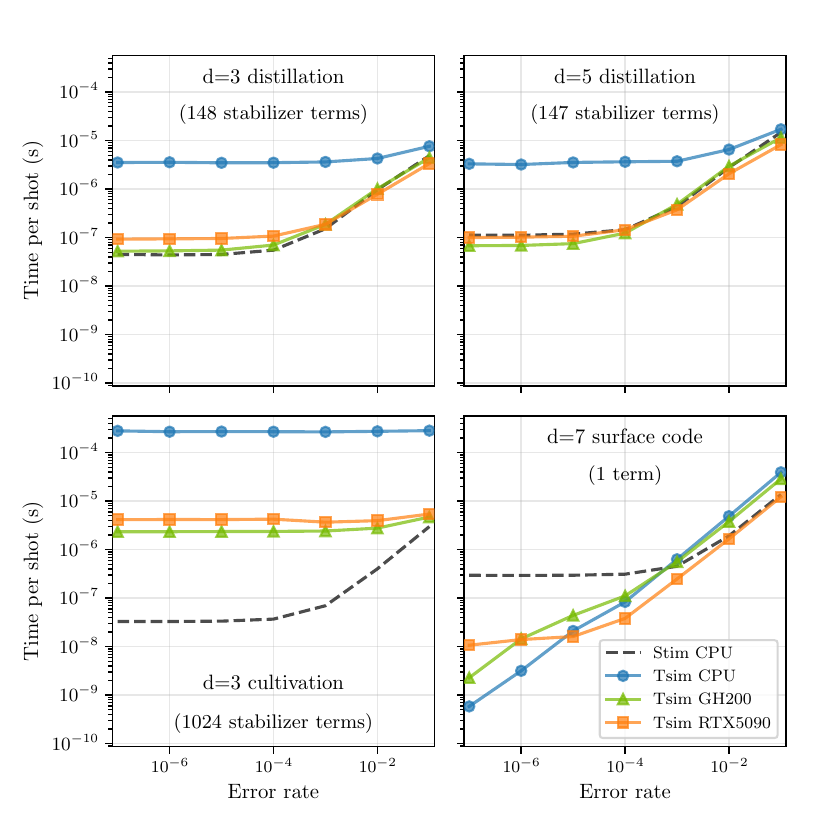}
  \caption{Shot-runtime comparison between Tsim and Stim for
    $d=3,5$ distillation circuits from \cite{sales2025experimental}, $d=3$ cultivation circuit from \cite{gidney2024magic}, and a $d=7$ rotated surface code. All CPU benchmarks (blue dots) were performed on an Apple M4 Pro. GPU benchmarks were conducted on an NVIDIA Grace Hopper GH200 system with a 72-core ARM Neoverse-V2 CPU (green triangles) or an NVIDIA RTX 5090 GPU with AMD Ryzen 9 9900X (orange squares).
  }
  \label{fig:benchmark-unnormalized}
\end{figure}

Figure~\ref{fig:benchmark}(a) shows normalized runtime comparison between Tsim and Stim, where Tsim runtime is normalized by the number of stabilizer terms in the decomposition. For reference, we present unnormalized data in Figure~\ref{fig:benchmark-unnormalized}.

Figure~\ref{fig:benchmark-unnormalized} additionally shows execution time for both GH200 and RTX 5090. Here, Tsim throughput is marginally higher on the GH200 system.

\section{Related simulators}

In Table~\ref{tab:simulator_comparison}, we present a high-level comparison of Tsim with two related simulators: Stim and SOFT~\cite{li2025soft}. In spirit, Tsim is related to Stim's Flip Simulator, whereas SOFT is closer to Stim's Tableau Simulator. This is because Tsim follows Stim's compile-once, sample-many pattern, while SOFT is a GPU-accelerated extended tableau simulator that supports non-Clifford gates. In the low-magic limit, Tsim has linear asymptotic scaling in the number of gates, while SOFT and Stim's Tableau Simulator scale at least quadratically in the number of qubits.

Reference~\cite{li2025soft} reports a sampling time of $6.68\,\mu$s per shot for the $d=3$ magic state cultivation circuit~\cite{gidney2024magic} on a single H800 GPU, similar to Tsim's $2.4\,\mu$s on GH200. However, SOFT scales to the $d=5$ cultivation circuit, which Tsim is not yet able to simulate.

A different simulation strategy has been introduced in \cite{surti2025efficient}. Here, efficient simulation relies on a low-rank decomposition of the noiseless gadget and the fact that, for many gadgets, Pauli errors propagate to Clifford errors at the end of the circuit. The authors report shot times $>10$ms for distance-3 cultivation circuit. Note that this approach is restricted to circuits that propagate Pauli errors to Cliffords.

\begin{table}[t]
  \centering
  \small
  \setlength{\tabcolsep}{4pt}
  \renewcommand{\arraystretch}{1.15}
  \resizebox{\linewidth}{!}{%
  \begin{tabular}{@{}lllllll@{}}
    \toprule
    Name & Type & Universal & Compilation & Sampling (per shot) & GPU-accel. & SIMD \\
    \midrule
    Tsim & Stabilizer rank & Yes & $\bigO{\ngates^3 + 2^{\alpha \nmagic} \nmagic^2}$ & $\bigO{\perr \ngates \navgflip + 2^{\alpha \nmagic} \nmagic^2 }$ & Yes (JAX) & Yes (JAX) \\
    Stim (flip)~\cite{gidney2021stim} & Pauli frame / flips & Clifford & $\bigO{\numqubits \ngates + \numqubits^3}$ & $\bigO{\ngates}$ & No & Yes \\
    Stim (tableau)~\cite{gidney2021stim} & Tableau & Clifford & none & $\bigO{\numqubits \ngates + \numqubits^3}$ & No & Yes \\
    SOFT~\cite{yoder2012generalization, li2025soft} & Generalized tableau & Yes & none & $\bigO{2^{ \nmagic}\numqubits \nmagic + \numqubits^2 \ngates}$ & Yes (CUDA) & GPU only \\
    \bottomrule
  \end{tabular}%
  }
  \caption{High-level comparison of Tsim with several related simulators. Asymptotic costs are written using the notation of this paper and are intended only as qualitative comparisons across methods.}
  \label{tab:simulator_comparison}
\end{table}

\section{Tsim API}
\label{sec:api}

\subsection{Frontend}
\label{sec:frontend}

The Tsim frontend follows the Stim API and accepts Stim-style intermediate representations, as shown
in Listing~\ref{lst:detector_sampling}.
Additionally, Tsim is tightly integrated into the Bloqade framework \cite{bloqade2025}.
Listing~\ref{lst:tsim_frontend_example} shows a simple example to construct a Tsim circuit from a Bloqade kernel.

\begin{lstlisting}[caption={Constructing a Tsim circuit from a Bloqade kernel}, label={lst:tsim_frontend_example}]
from bloqade.tsim import Circuit

@squin.kernel
def main():
    q = squin.qalloc(2)
    squin.h(q[0])
    squin.t(q[0])
    squin.broadcast.depolarize(0.01, q)
    squin.cx(q[0], q[1])
    bits = squin.broadcast.measure(q)
    squin.set_detector(bits, coordinates=[0, 1])
    squin.set_observable([bits[0]], idx=0)

circuit = Circuit(main)
sampler = circuit.compile_sampler()
sampler.sample(shots=1_000_000, batch_size=100_000)
\end{lstlisting}
To sample detector and observable outcomes separately, one can use a detector sampler:

\begin{lstlisting}[caption={Sampling detector and observable outcomes separately.}]
sampler = circuit.compile_detector_sampler()
detector_bits, observable_bits = sampler.sample(
    shots=1_000_000,
    separate_observables=True,
)
\end{lstlisting}

\subsection{Compilation and sampling}

Tsim follows a compile-once, sample-many workflow. Constructing a sampler, especially with \texttt{compile\_detector\_sampler()}, can be expensive because it performs the underlying compilation step. The first call to \texttt{sample()} may also be slow due to JAX/XLA compilation overhead.

The \texttt{sample()} method takes two arguments: \texttt{shots} and \texttt{batch\_size}. By default, Tsim attempts to determine an appropriate batch size based on available RAM on CPU or VRAM on GPU. Changing \texttt{batch\_size} triggers JAX/XLA recompilation, so best performance is obtained by reusing the same batch size across repeated sampling calls.

\subsection{Visualization}

The \texttt{tsim.Circuit.diagram()} method supports several visualization modes.

The \texttt{timeline-svg} and \texttt{timeslice-svg} modes use Stim's SVG backend to display the circuit. The \texttt{pyzx} mode uses PyZX D3 plotting to show the unsimplified ZX diagram, including detector and observable annotations.

The \texttt{pyzx-meas} mode shows a full-reduced ZX diagram in which detectors and observables are removed and output vertices are connected directly to measurement outcomes. This graph forms the basis for autoregressive sampling, where marginal probabilities can be obtained by projecting or tracing over output vertices.

Finally, \texttt{pyzx-dets} shows a full-reduced ZX diagram in which output vertices are connected to detectors and observables. For deterministic Clifford circuits, this representation captures the Tanner graph of the quantum error-correcting circuit.
\end{document}